\documentclass[useAMS,usenatbib,aas_macros]{mn2e}
\input{psfig}
\usepackage{color}

\newcommand{\be}{\begin{equation}}
\newcommand{\ee}{\end{equation}}

\def\no{\noindent}

\newcommand{\ifm}[1]{\relax\ifmmode#1\else$\mathsurround=0pt #1$\fi}
\newcommand{\kms}{\ifmmode\,{\rm km}\,{\rm s}^{-1}\else km$\,$s$^{-1}$\fi}

\newcommand{\ltsima}{$\; \buildrel < \over \sim \;$}
\newcommand{\lsim}{\lower.5ex\hbox{\ltsima}}
\newcommand{\gtsima}{$\; \buildrel > \over \sim \;$}
\newcommand{\gsim}{\lower.5ex\hbox{\gtsima}}

\def\M11{M_{11}}
\def\V100{V_{100}}
\def\R1{R_{Mpc}}
\def\T6{T_6}

\title[Downsizing by Shutdown in Red Galaxies] 
{{Downsizing by Shutdown in Red Galaxies}} 

\author[A. Cattaneo, A. Dekel, S.~M. Faber, B.~Guiderdoni]
{A.~Cattaneo$^{1\star}$, A.~Dekel$^{2,\dagger}$, S.~M.~Faber$^{3}$, B.~Guiderdoni$^{4}$\\
\\
$^1$Astrophysikalisches Institut Potsdam, an der Sternwarte 16, 14482 Potsdam, Germany\\
$^2$Racah Institute of Physics, Hebrew University of Jerusalem, 91904 Jerusalem, Israel\\
$^3$UCO/Lick Observatory, University of California, USA\\
$^4$Centre de Recherche Astrophysique de Lyon, France\\
$^\star$acattaneo@aip.de, $^\dagger$dekel@phys.huji.ac.il }
\begin{document}

\pagerange{\pageref{firstpage}--\pageref{lastpage}} \pubyear{2005}

\maketitle

\label{firstpage}


\begin{abstract}

We address the origin of the `downsizing' of elliptical galaxies,
according to which the stars in more massive galaxies formed earlier
and over a shorter period than those in less massive galaxies.
We show that this could be the natural result of a shutdown of star formation 
in dark matter haloes above a critical mass of $\sim 10^{12} M_\odot$.
This is demonstrated using a semi-analytic simulation of galaxy formation
within the standard hierarchical scenario of structure formation.
The assumed threshold mass is motivated by the prediction of stable shock 
heating above this mass and the finding that such a shutdown reproduces 
the observed distribution of galaxies in luminosity and colour.
The shutdown at a critical halo mass introduces a characteristic stellar mass
for the transition of
galaxies into the `red sequence' of the galaxy colour-magnitude diagram.
Central galaxies of haloes that are more massive today have reached this mass
earlier and can therefore grow further along the red sequence by dry mergers, 
ending up more massive and containing older stars.
Small galaxies formed in haloes below the critical mass can shutdown
late, when they fall into haloes above the critical mass and become satellites.
While our semi-analytic simulation that incorporates an explicit shutdown 
reproduces downsizing as inferred from the stellar ages of ellipticals, 
we explain why it is much harder to detect downsizing using the mass 
functions of different galaxy types.
\end{abstract}



\section{Introduction}
\label{sec:intro}

The term `downsizing' was coined by \citet{cowie_etal96} to describe the 
decline with time of the $K$-band rest-frame luminosity of the galaxies 
with the highest specific star formation rate 
$\dot{M}_{\rm star}/M_{\rm star}$ (SFR)
as observed in the redshift interval $0.2<z<1.7$. 
This `downsizing in time' has been confirmed by later studies 
\citep{guzman_etal97,brinchmann_ellis00,kodama_etal04,juneau_etal05,
bell_etal05,noeske_etal07}.
It can be viewed as part of a more general phenomenon valid robustly across 
the Hubble sequence of galaxy types, in the sense that galaxies of later type,
which are typically less massive, are known to have formed their stars most 
efficiently at later times and over longer periods 
\citep{searle_etal73,tinsley73,sandage86}.
A possibly related downsizing is evident in the accretion histories 
of supermassive black holes \citep{steffen_etal03,ueda_etal03,heckman_etal04,
barger_etal05,hasinger_etal05}. 
 
Another form of downsizing, termed `archaeological downsizing' \citep{thomas_etal05}, is inferred 
from the stellar populations of today's galaxies. 
Star formation histories derived from observed line indices and abundance 
ratios using stellar evolution models reveal a strong correlation between 
mean stellar age and galactic stellar mass both in elliptical galaxies
\citep{nelan_etal05,thomas_etal05,graves_etal07} and in the general
population of galaxies from the large Sloan Digital Sky Survey
(SDSS; \citealp{heavens_etal04}; \citealp{jimenez_etal05}; \citealp{panter_etal06}). 
In the current study we focus on archaeological downsizing, as we address 
red galaxies that show only little active star formation at present.

At a first glance, the observed downsizing seems to be in conflict with
the standard wisdom of hierarchical structure formation, according to which
smaller objects collapse earlier and gradually assemble into more massive 
objects.
This apparent conflict is relaxed when realizing that 
star formation and gravitational assembly are two distinct processes.
In particular, stars can form first in the small building blocks of 
today's massive galaxies without violating the assembly hierarchy of each
galaxy.  
In fact, if gas processes limit galaxy formation to dark matter haloes
above a minimum mass, a certain downsizing trend arises naturally from
the hierarchical dark matter assembly process itself \citep{neistein_etal06}.
Still, the origin of downsizing as observed requires a more quantitative
theoretical understanding.

Recent developments in the modelling of galaxy formation 
\citep[e.g.][]{cattaneo_etal06,bower_etal06,croton_etal06}
have been driven by the realization that galaxies are divided into 
two major distinct types: blue star-forming late-type galaxies in low-density
environments, red and dead early-type galaxies in groups and clusters
(\citealp{lin_etal97,lin_etal99,im_etal01,strateva_etal01,kauffmann_etal03b,
baldry_etal04,balogh_etal04,bell_etal04,hogg_etal04,weiner_etal05};
see \citealp{dekel_birnboim06} for a summary of the bimodality phenomenon).
Combined with the fact that blue galaxies are confined below
a critical stellar mass of $M_{\rm star}^{\rm crit}\sim 3\times 10^{10}M_\odot$ 
while the population above this mass is dominated by red galaxies,
the colour bimodality itself is evidence for downsizing. 
It indicates that the most massive galaxies, the red ones, have converted 
their gas into stars several billion years ago, while less massive 
galaxies, which are mostly blue, are still making stars.

Downsizing is not only a reflection of the galaxy bimodality.  
It is also valid separately within each of the two populations,
the red one \citep{nelan_etal05,thomas_etal05,graves_etal07} and the blue one 
\citep{drory_etal06,noeske_etal07}. Less massive galaxies contain
younger stars even within the same spectral type.
Nevertheless, we are tempted to consider the possibility that downsizing
and bimodality originate from the same underlying physics.

The existence of a characteristic mass for massive galaxies emerges from the
the competition between the gravitational dynamical time and the radiative 
cooling time \citep{rees_ostriker77,binney77,silk77,white_rees78,
blumenthal_etal84}. 
\citet{birnboim_dekel03} and \citet{dekel_birnboim06} showed that a stable
shock can expand to the virial radius and be supported against gravity 
only in haloes above a critical mass of $M_{\rm shock}\sim 10^{12}M_\odot$,
producing a hot medium at the virial temperature of $T \geq 10^6$K. 
In smaller haloes, efficient cooling does not allow an extended
stable shock, so the accreting gas flows cold into the centre, leading 
to the buildup of discs and efficient star formation.
This phenomenon was detected in parallel in cosmological simulations
\citep{keres_etal05,birnboim_etal07,cattaneo_etal07}, 
confirming the existence of a threshold mass that is quite
independent of redshift. 
\citet{dekel_birnboim06} argued that the transition from cold flows to the 
shutdown of gas supply and star formation at $M_{\rm shock}$, which is
predicted to be more pronounced after $z \sim 2-3$, naturally leads to many 
of the observed features associated with the galaxy bimodality. In particular,  
the bimodality scale $M_{\rm star}^{\rm crit}$ is the typical stellar mass of 
the central galaxy in a halo with mass $M_{\rm shock}$. 
The implications of such a scenario have been studied using semi-analytic
simulations by \citet{cattaneo_etal06}, \citet{croton_etal06} and 
\citet{bower_etal06}, each applying a slightly different shutdown procedure
driven by halo mass (plus, sometimes, bulge mass or central black hole mass).
The formation of the red sequence through the quenching and reddening of blue galaxies has also been
studied by \citet{bell_etal04} and \citet{faber_etal07}.
Here we highlight a particular aspect of the quenching mechanism, 
associated with the fact that the critical mass for quenching is roughly 
constant after $z \sim 2-3$.

In this article we demonstrate that the shutdown of star formation above a 
critical halo mass $M_{\rm crit}\sim M_{\rm shock}$ at $z < 3$ 
accounts for downsizing in the red galaxy population.
The key idea is that central galaxies stop making stars and enter the red sequence at a critical 
stellar mass $M_{\rm star}^{\rm crit}$ of the order of the baryonic mass in a halo of mass $M_{\rm shock}$.
Galaxies with higher final stellar mass reach $M_{\rm star}^{\rm crit}$ 
earlier and therefore have more time to grow through dissipationless (`dry') 
mergers along the red sequence.  

There are three complications to this simple picture, which we discuss in this article.
First, satellite galaxies shut down earlier than central galaxies with the same mass, and they often do it 
before they have reached $M_{\rm star}^{\rm crit}$ (e.g. because their halo has merged into another one with mass $>M_{\rm shock}$).
Second, the expected penetration of narrow cold streams through the hot media of massive high-redshift haloes 
elevates the effective $M_{\rm crit}$ to values above $M_{\rm shock}$ at $z\geq 3$ \citep{dekel_birnboim06,cattaneo_etal06}. 
Third, additional shutdown mechanisms may be at work, e.g. the exhaustion of gas after
major mergers and/or feedback from the central black hole.
We have attempted to mimick these effects by intoducing a further shutdown criterion related to the bulge-to-disc ratio. 
The inclusion of a bulge-to-disc ratio criterion as a sufficient condition for quenching
in addition to the halo-mass criterion improves the fit to the joint colour-magnitude distribution of galaxies by accounting for
the presence of red galaxies in haloes below $M_{\rm crit}$,
but its overall effect is fairly minor (\citealp{cattaneo_etal06}, Fig.~9).

Downsizing has also been involved to explain the way the
stellar-mass functions of galaxies of different colors or morphological 
types evolve with time.  
In particular, while the comoving number density of blue galaxies has not
changed much since $z \sim 1$, the number of red galaxies has seemingly grown 
by a larger factor at the faint end than at the bright end
(\citealp{rudnick_etal03,bell_etal04,drory_etal04,bundy_etal05,
drory_etal05,borch_etal06,pannella_etal06,faber_etal07}).
We argue below that the interpretation of this behavior as downsizing is 
hampered by severe measurement uncertainties (see the Appendix) and conceptual misunderstandings.
  
The paper outline is as follows.
In \S~2, we describe our model of galaxy formation explored by means of a semi-analytic modelling 
technique applied to merger trees from a cosmological N-body simulation, following 
 \citet{cattaneo_etal06}. 
In \S~3, we elaborate on how the shutdown of star formation leads to downsizing.
In \S~4, we show how the shutdown produces downsizing in the buildup of the red sequence,
in the sense that the high-mass end is populated at earlier times.
In \S~5, we perform a comparison with the archeological downsizing inferred from the stellar ages
in elliptical galaxies \citep{thomas_etal05}.
In \S~6, we consider the characteristic times of galaxy formation in terms of star formation history and mass assembly history, 
and use this analysis to address the apparent conflict between downsizing and the hierarchical model.
In \S~7, we explain how downsizing in star formation and upsizing in mass assembly combine to produce the 
evolution of the mass functions of red and blue galaxies, and why
it is hard to detect downsizing using the mass 
functions of different galaxy types.
In \S~8, we summarize the conclusions of the article.
The downsizing in blue galaxies will be addressed in a separate publication.

\section{The galaxy formation model} 
\label{sec:galics} 

GalICS (Galaxies In Cosmological Simulations; \citealp{hatton_etal03}) 
is a method to simulate the formation of galaxies in a $\Lambda$CDM Universe.
It combines high-resolution cosmological N-body simulations of the 
gravitational clustering of the dark matter with a semi-analytic (SAM) approach 
to the physics of the baryons (gas accretion, galaxy mergers, star formation 
and feedback).  The version of GalICS used here, as well as the adopted
values of the parameters, are the same as described in \citet[][and references
therein]{cattaneo_etal06}.  We summarize here the relevant issues.

\subsection{Dark-matter simulation}

The cosmological N-body simulation that follows the hierarchical clustering 
of the dark-matter component has been carried out with the parallel tree 
code.
The assumed cosmological model is a flat $\Lambda$CDM Universe with 
a cosmological constant of $\Omega_{\Lambda}=0.667$, 
a Hubble constant of $H_0=66.7{\rm\,km\,s}^{-1}$,
and a $\Lambda$CDM power spectrum of initial fluctuations normalized 
to $\sigma_8 =0.88$.
The computational volume is a cube of side $(150{\rm\,Mpc})^3$ with $256^3$ 
particles of $8.3\times 10^9$M$_{\odot}$ each
and a smoothing length of $29.3\,$kpc. 
The simulation produced 100 output snapshots spaced logarithmically in the 
expansion factor $(1+z)^{-1}$ from $z=35.59$ to $z=0$.

We have analysed each snapshot with a friend-of-friend algorithm 
\citep{davis_etal85} to identify 
virialized haloes containing more than 20 particles.
The minimum halo mass is thus $1.65\times 10^{11}M_\odot$.
The three global properties characterizing each halo to be used in the 
SAM are the virial mass, the virial density, and the spin 
parameter.
Merger trees are constructed by linking the haloes identified in each snapshot 
with their progenitors in the previous snapshot, that is,
all predecessors from which the halo has inherited one or more 
particles.
We do not use substructure information from the N-body simulation. 
Instead, once a halo becomes a subhalo of another halo, we switch from 
following its evolution with the N-body integrator to an approximate 
treatment based on semi-analytic prescriptions.          

\subsection{Semi-analytic modeling of gas processes}

A newly identified halo is assigned a gas mass based on a universal 
baryon fraction $\Omega_{\rm b}/\Omega_0 = 0.135$. 
The conditions for the formation of a galaxy at its centre are that
the halo is gravitationally bound and that its angular momentum parameter 
is $\lambda<0.5$.
We have simulated a `standard' and a `new' model, which differ
in the way gas from the halo accretes onto the central galaxy. 

In the `standard' model, the initial gas distribution is a singular 
isothermal sphere, at the virial temperature, truncated at the virial radius.
The cooling time of the hot gas is computed using
the radiative cooling function of \citet{sutherland_dopita93}.
The gas for which both the cooling time and the free fall time are shorter
than the timestep $\Delta t$ between two N-body snapshots is accreted 
onto the central galaxy during that time interval. The transfer of baryons 
from the halo to the galaxy is accompanied by an inflow of hot gas
to keep the hot gas distribution a truncated singular isothermal sphere. 
The numerical accuracy and the self-consistency of this cooling algorithm 
have been verified by comparison with a cosmological hydrodynamic simulation 
based on the same physics and the same dark-matter realization 
\citep{cattaneo_etal07}.

\begin{figure*} 
\noindent
\label{mhalo_mgal}
\begin{minipage}{8.4cm}
  \centerline{\hbox{
      \psfig{figure=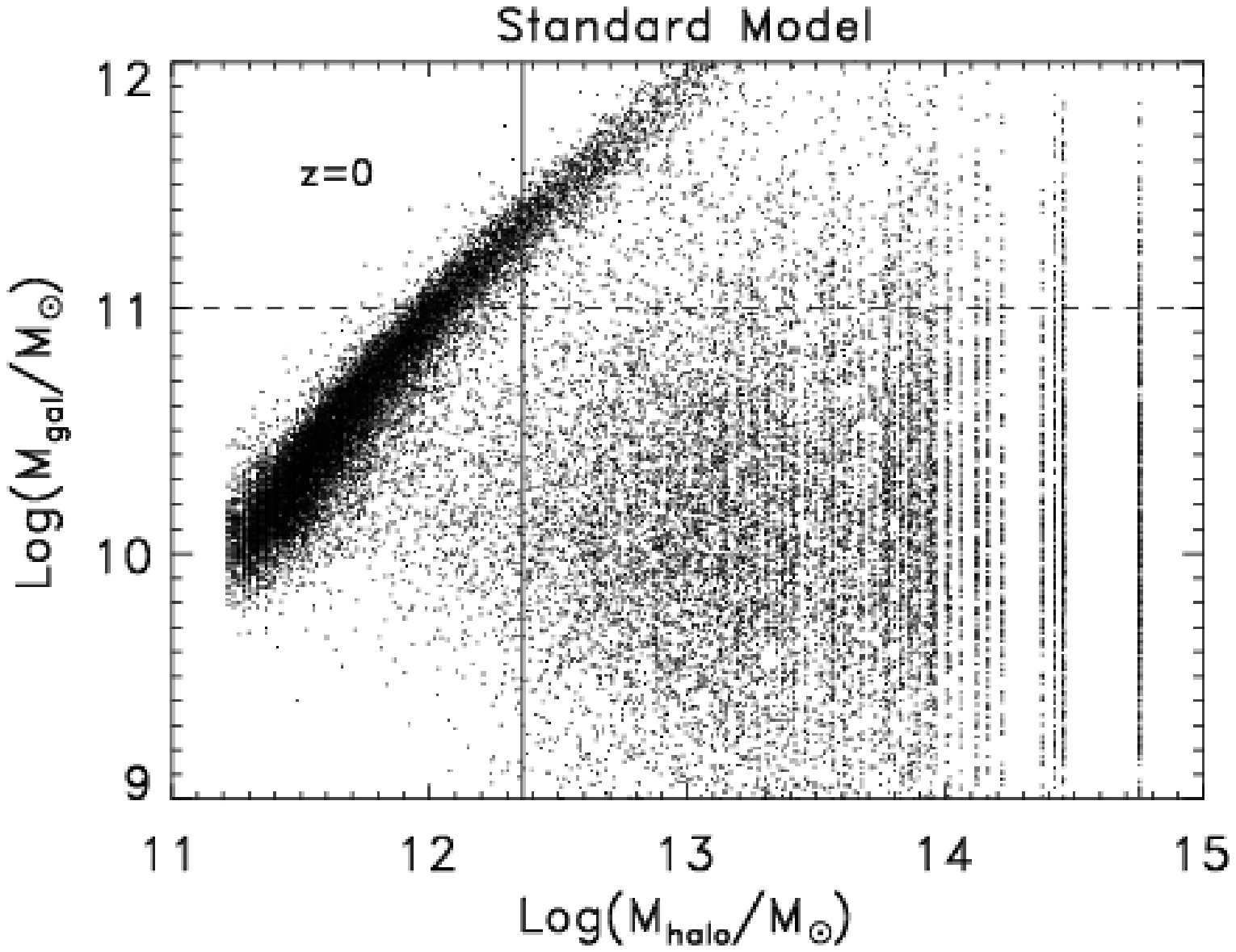,height=5.73cm,angle=0}
  }}
\end{minipage}\    \
\begin{minipage}{8.4cm}
  \centerline{\hbox{
      \psfig{figure=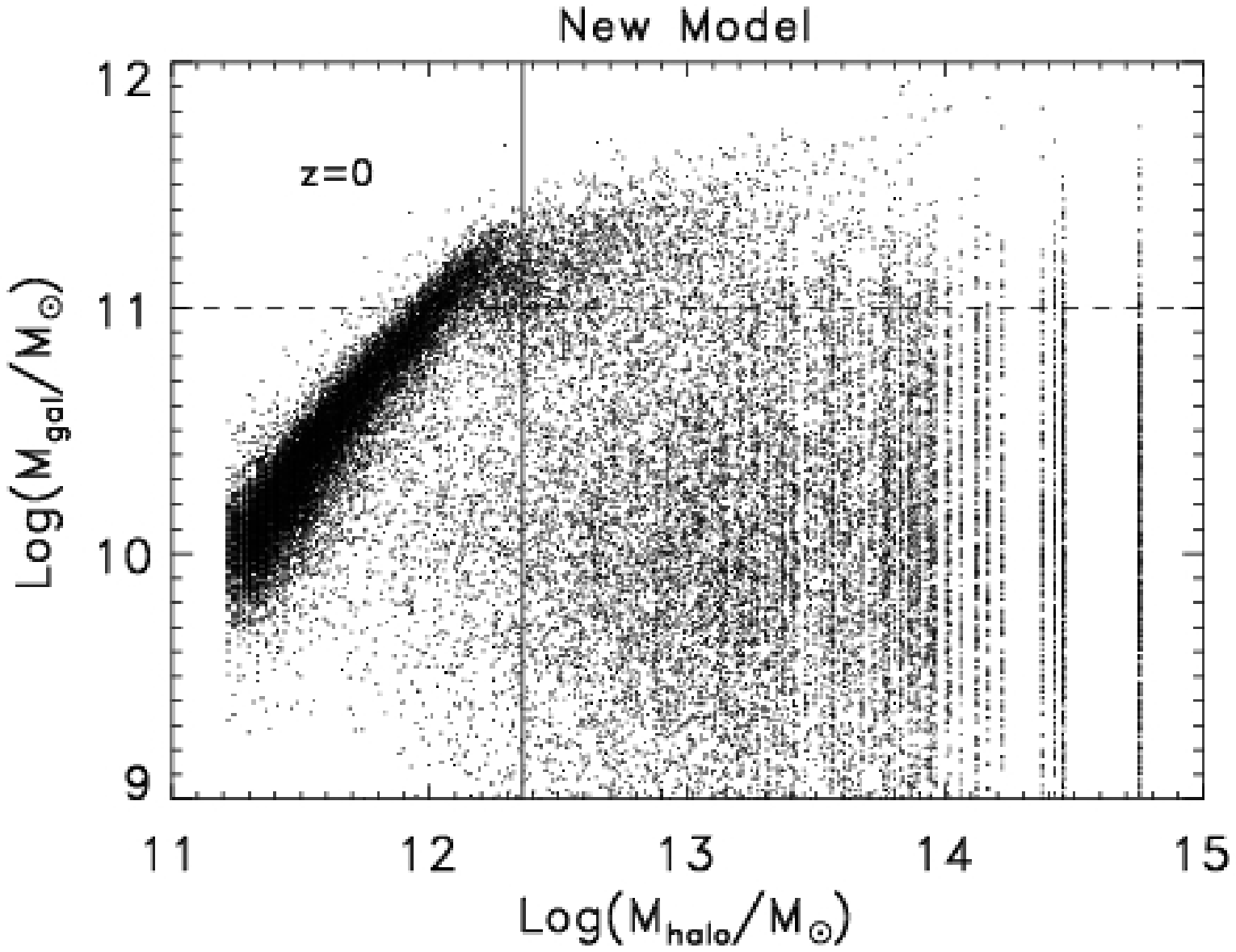,height=5.73cm,angle=0}
  }}
\end{minipage}\    \
\caption{Galaxy stellar mass versus halo mass in the standard model (left)
and the new model with shutdown (right). In the new model, the general
growth of $M_{\rm star}$ with $M_{\rm halo}$ breaks off at 
$M_{\rm halo}\sim M_{\rm shock}\sim 2\times 10^{12}M_\odot$.
Differences at $M_{\rm halo}< M_{\rm shock}$ are due to the additional
shutdown of accretion onto the galaxies that have become bulge-dominated.
Objects with high $M_{\rm halo}$ but low $M_{\rm gal}$ are mostly satellites that have fallen into larger haloes.}
\label{mgalvsmhalo}
\end{figure*}

In the `new' model, we introduce the proposed quenching above a threshold 
halo mass, following \citet{dekel_birnboim06}.
Once a halo grows above $M_{\rm shock}\sim 10^{12}M_\odot$,
the halo gas is assumed to be shock-heated to the virial temperature.
Unlike previous models, once heated, the hot gas is kept hot forever.
This can be due to several different long-term quenching mechanisms,
such as self-regulated radio-AGN feedback  
\citep[e.g.][and references therein]{croton_etal06,cattaneo_teyssier07},
shocked accretion \citep{birnboim_etal07},
or gravitational quenching by clumpy accretion
\citep{dekel_birnboim07,khochfar_ostriker07}. 
The critical mass for quenching is assumed to be
\begin{equation}
M_{\rm crit}=M_{\rm shock}\times {\rm max}\{1,\,10^{1.3(z-z_{\rm c})}\} \ ,
\label{eq:mcrit} 
\end{equation}
where $M_{\rm shock}\sim 10^{12}M_\odot$ is the shock-heating scale 
and the term $\propto 10^{1.3z}$ accounts for the
penetration of cold streams in massive haloes at high redshift
\citep[][Fig.~7]{dekel_birnboim06}.
In \citet{cattaneo_etal06}, we set the parameters to 
$M_{\rm shock}= 2\times 10^{12}M_\odot$ and $z_{\rm c}=3.2$ 
by optimizing the way the new model fits the colour-magnitude distribution 
at $z\simeq 0$ and the Lyman-break galaxy luminosity function at $z\simeq 3$. 

In the standard model, there is no explicit shutdown of cooling, and the mass and luminosity of the central galaxy grow roughly
linearly with halo mass over the entire range resolved by the N-body
simulation. Shutting down the hot accretion mode when 
$M_{\rm halo}>M_{\rm crit}$ introduces a characteristic mass 
(Fig.~\ref{mgalvsmhalo}) and luminosity, which mark the separation between  
blue and red galaxies.

Initially, all galaxies are assumed to form at a distance $r_i=0$ 
from the centre of their halo.  When two or more haloes merge, 
their galaxies are repositioned at a distance $r_f$ from the
centre of the new halo, determined such that
$r_f\rightarrow r_i$ for $M_f/M_i\rightarrow 1$ and 
$r_f\rightarrow r_{\rm vir}$ for $M_f/M_i\rightarrow\infty$
(where $M_i$ and $M_f$ are the halo mass before and after the merger
and $r_{\rm vir}$ is the virial radius of the merged halo.
Halo mergers rapidly create haloes with more than one galaxy,
which we identify as galaxy groups and clusters. However, only the 
central galaxy is allowed to accrete gas from the halo (prior to quenching).

\subsection{Morphologies}

GalICS models a galaxy with three components: a disc, a bulge and a 
transitional `starburst' component. 
Each component may contain stars and cold gas,
while the hot gas is treated as a component of the halo.
Only the disc of the central galaxy accretes gas from the halo. 
The bulge grows by mergers and by disc instabilities.
While stars are trasferred directly from the disc to the bulge, the gas in transit from the disc to the bulge is assumed 
to pass through an intermediate starburst
component, where its star formation timescale decreases by a factor of ten with respect to 
that of the bulge (Section~2.4).
Stars formed in the starburst are moved to the bulge 
after they have reached an age of $100\,$Myr. 
The only gas in the bulge is that from stellar mass loss (\S~2.4).

The disc profile is assumed to be exponential, with its radius determined by 
conservation of angular momentum. 
The bulge is assumed to have a \citet{Hernquist90} profile and its 
radius is determined based on an energy conservation argument.

We now elaborate on the two mechanisms for transferring baryons from the disc to the spheroid.
First, disc instabilities transfer matter from the disc to the spheroid (stars to the bulge, gas to the starburst) until the 
bulge is massive enough to stabilize the disc.
The stability criterion is $v_{\rm rot}<0.7v_{\rm c}$, where $v_{\rm rot}$ 
and $v_{\rm c}$ are the disc's rotation and circular velocity at the disc's 
half mass radius \citep{vandenbosch98}. 
Second, dynamical friction, modelled as in \citet{hatton_etal03},
drives galaxies to the centre of their dark matter halo 
and is the dominant cause of galaxy merging, although we
also include satellite-satellite encounters. 
The fraction of the disc that is transferred to the bulge in a merger grows 
with the mass ratio of the merging galaxies. It ranges from zero
for a very minor merger to unity for an equal mass merger. 
This simplified picture of the dynamics of morphological 
transformations provides results consistent with the key observational
constraints such as the Faber-Jackson relation and the Fundamental Plane of 
spheroids \citep{hatton_etal03}.  

\begin{figure*} 
\noindent
\begin{minipage}{5.5cm}
  \centerline{\hbox{
      \psfig{figure=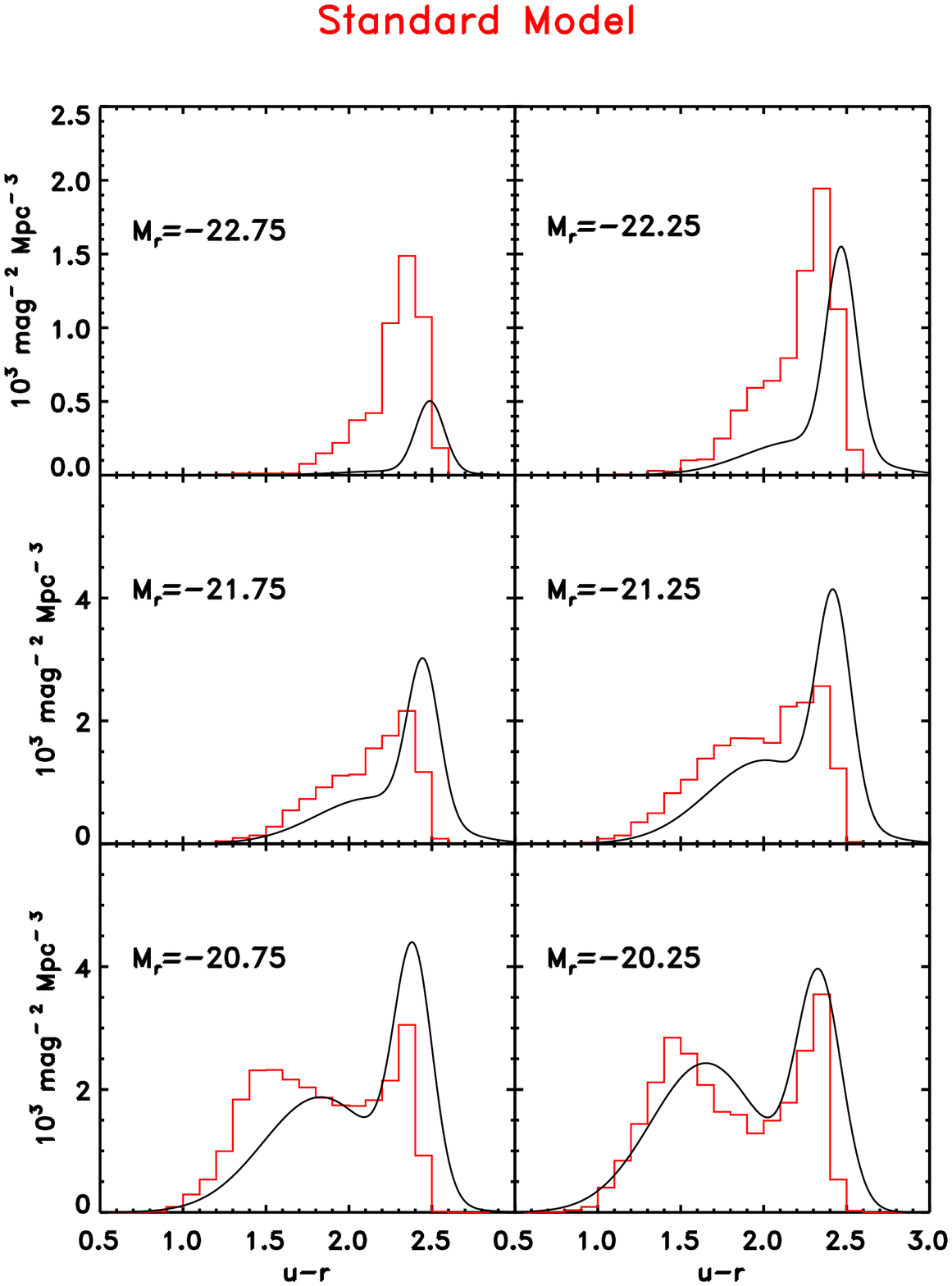,height=8.cm,angle=0}
  }}
\end{minipage}\    \
\noindent
\begin{minipage}{5.5cm}
  \centerline{\hbox{
      \psfig{figure=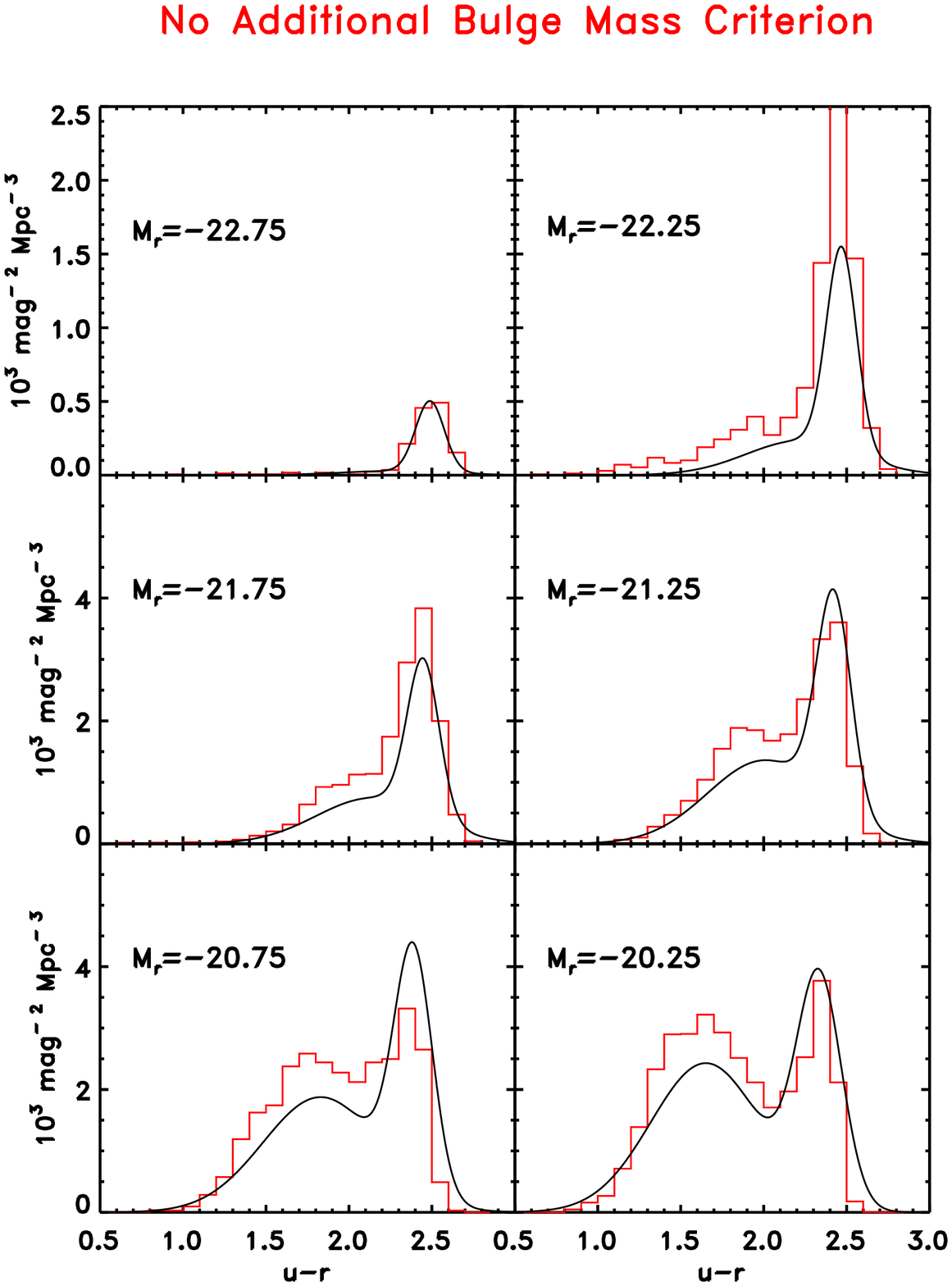,height=8.cm,angle=0}
  }}
\end{minipage}\    \
\begin{minipage}{5.5cm}
\centerline{\hbox{
\psfig{figure=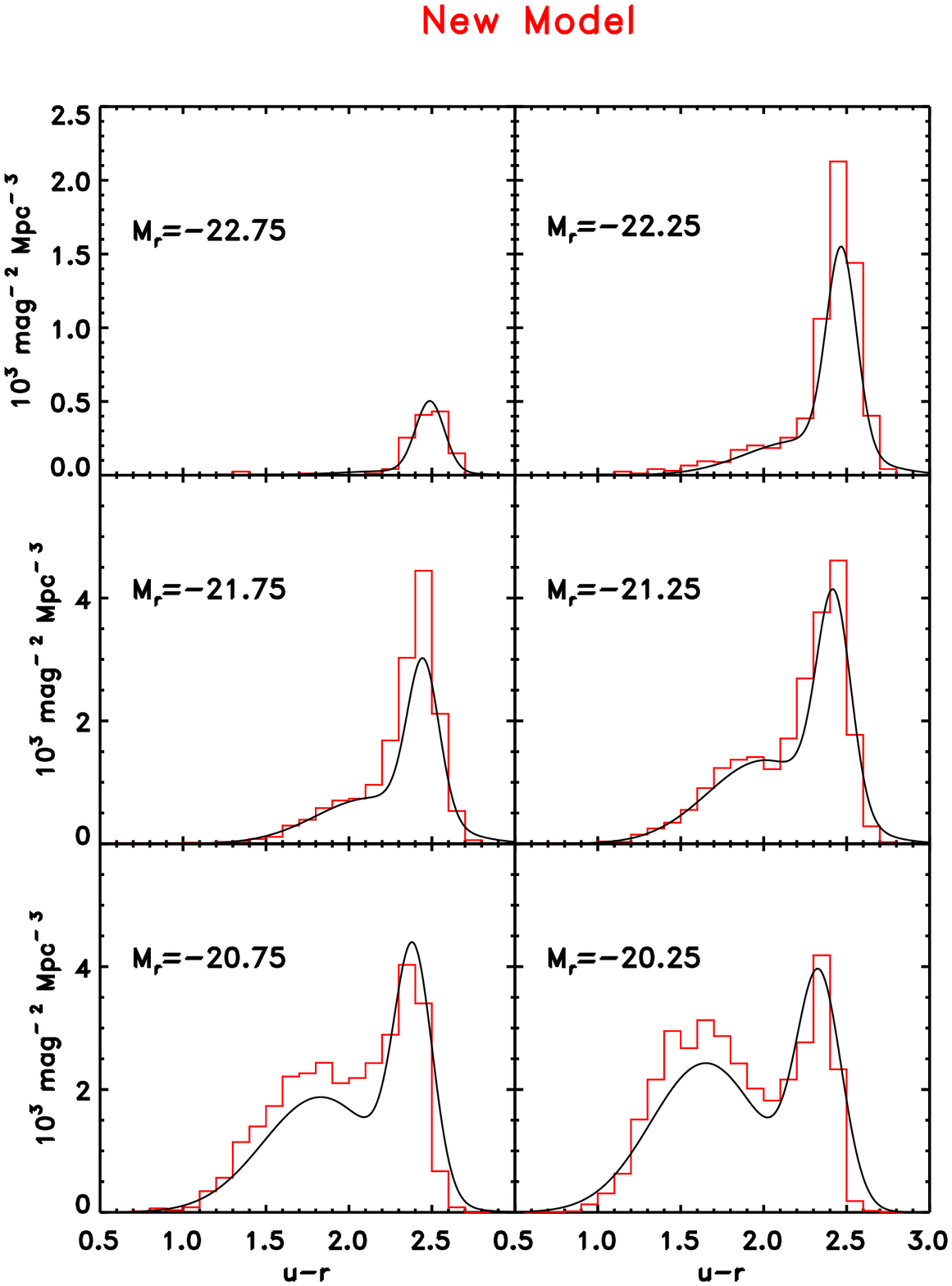,height=8.cm,angle=0}
  }}
\end{minipage}\    \
\vskip 0.3cm
\caption{The the $u-r$ colour distribution in $r$-band magnitude bins 
for the SDSS data (\citealp{baldry_etal04};
black smoothed histogram) and for three different versions of our SAM
(red curves).
Left: the standard model without an explicit shutdown of central galaxies
but with a standard shutdown of gas accretion onto satellites.
Middle: A model with shutdown of star formation once 
$M_{\rm halo}>M_{\rm crit}$. 
Right: Same as in the middle, with the additional shutdown when
$M_{\rm bulge}>M_{\rm star}/2$ (reproduced from \citealp{cattaneo_etal06}).}
\label{colour_histograms}
\end{figure*}

\subsection{Star formation and feedback}

The star formation law is the same for all components: 
\begin{equation}
\label{eq:sfr} 
\dot{M}_{\rm star}={M_{\rm cold}\over\beta_*t_{\rm dyn}}(1+z)^{\alpha_*} \ .
\end{equation}
The mass of cold gas, $M_{\rm cold}$, refers to the component in question,
and $t_{\rm dyn}$ is the dynamical time 
(corresponding to half a rotation 
for discs and half a crossing time for bulges).
In the standard GalICS model, 
star formation is activated when the gas surface density is
$\Sigma_{\rm gas}>20\,m_{\rm p}{\rm\,cm}^{-2}$ 
($m_{\rm p}$ is the proton mass).
Furthermore, by imposing a minimum halo mass of 
$M_{\rm halo}=1.65\times 10^{11}M_\odot$ 
due to the N-body resolution, we effectively assume
that there is no star formation in haloes below this mass,
potentially mimicking the effects of more aggressive supernova feedback in small haloes (see below). 
Moreover, in the new model we assume a complete shutdown of star formation in haloes with mass $>M_{\rm crit}$
by removing all the cold present in any component of these galaxies and by moving this gas to the halo's hot component.

The star formation efficiency parameter is assumed to be
$\beta_*=50$ \citep{guiderdoni_etal98}, the same for all components.
We have $\alpha_*=0$ in the standard GalICS model but allow an
enhanced star-formation rate at high redshift by adopting $\alpha_*=0.6$ 
in the new  model, to mimic the rapid accretion by cold flows.
This improves the fit to the luminosity function of Lyman-break galaxies 
\citep{cattaneo_etal06}.

Gas in transit from the disc to the bulge, whether by mergers or disc instabilities, passes through a starburst phase 
in which the SFR grows by a factor of 10 based on the dynamical time of the bulge.
This high SFR is obtained 
by assuming that the starburst 
radius is ten times smaller than the final bulge radius 
(see the hydrodynamic simulation in \citealp{cattaneo_etal05a}). 
As we anticipated when we described the different components of a galaxy, the stars formed in the starburst component stay in the starburst 
component for $100\,$Myr  and then are moved to the bulge.
The properties of low redshift galaxies are insensitive 
to the starburst star formation timescale as long as it is much shorter 
than the Hubble time.

We assume a \citet{kennicutt83} initial mass function.
Stars are evolved between snapshots using substeps of at most 1\,Myr. 
During each sub-step, stars release mass and energy into the interstellar 
medium. Most of the mass comes from the red giant and 
the asymptotic giant branches of stellar evolution, while most of the energy 
comes from shocks due to supernova explosions. The enriched material 
released in the late stages of stellar evolution is mixed with 
the cold phase, while the energy released from supernovae is used to reheat 
the cold gas and to return it to the hot phase in the halo. 
Reheated gas is ejected from the halo if the potential is shallow enough.
The rate of mass loss through supernova-driven winds $\dot{M}_{\rm w}$ 
is determined by the equation
\begin{equation} 
{1\over2}\dot{M}_{\rm w}v_{\rm esc}^2
=\epsilon_{\rm SN}\eta_{\rm SN}E_{\rm SN}\dot{M}_{\rm star},
\label{eq:mdot}
\end{equation}
where $E_{\rm SN}=10^{51}\,$erg is the energy of a supernova, 
$\eta_{\rm SN}=0.0093$ is the number of supernovae for
$1\,M_\odot$ of stars formed and $v_{\rm esc}$ is the escape velocity 
\citep{dekel_silk86}.
In GalICS, we use $v_{\rm esc}\simeq 1.84v_{\rm c}$ for discs and 
$v_{\rm esc}=2\sigma$ for bulges and starbursts.
The supernova efficiency $\epsilon_{\rm SN}\simeq 0.2$ is similar to those 
commonly adopted in SAMs \citep{somerville_primack99,cole_etal00}. 

\subsection{Stardust}
  
The STARDUST model \citep{devriendt_etal99} provides the spectrum of a stellar 
population as a function of age and metallicity and includes a
phenomenological treatment of the reprocessing of stellar light by dust.
GalICS uses STARDUST to compute the spectrum of each component and 
to output galaxy magnitudes.

Dust absorption is computed with a phenomenological extinction law that
depends on the column density of neutral hydrogen, the metallicity of the 
obscuring material and the randomly selected inclination angle (for spirals).
The re-emitted spectrum is the sum of four templates (big and small carbon 
grains, silicates, and polycyclic aromatic hydrocarbons). 
Their relative weights are chosen to reproduce the relation between 
bolometric luminosity and infrared colours observed locally in IRAS galaxies.
In GalICS we only consider each component's self-absorption, e.g. the dust in a starburst only obscures
the starburst's light, not the old bulge stellar population.

\begin{figure*} 
\noindent
\centerline{\hbox{
            \psfig{figure=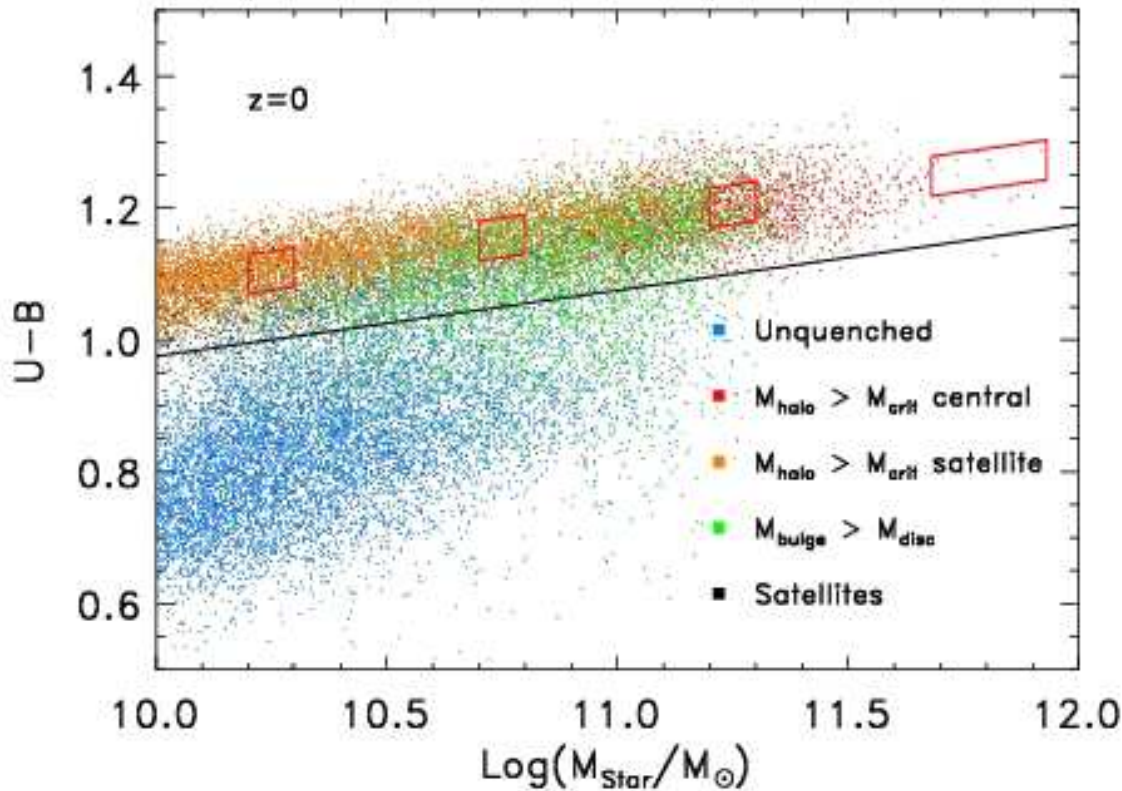,height=12.cm,angle=0}
  }}
\caption{Colour-mass diagram for the new model with shutdown.
The `green valley' separating the red sequence from the blue cloud is marked
by the black line.
Blue symbols mark galaxies in which the SFR has not been explicitly shut down.
Red symbols denote central galaxies in which the SFR has been shut down 
because their halo has grown above the critical mass. 
Orange symbols show the population of satellite galaxies in haloes above the critical mass.
Green symbols are central galaxies of haloes with 
$M_{\rm halo}<M_{\rm crit}$, where
the formation of stars has been shut down because the bulge has become 
the dominant component. 
Black symbols refer to galaxies that do not accrete gas simply because they are
satellites. 
In this diagram galaxies have been coloured according to their status at $z=0$, and not according
to the first mechanism that has intervened and caused the shutdown of star formation in the history of a given galaxy.
The red parallelograms show the four bins along the red sequence 
from which we have selected the galaxies for
Figs.~\ref{halo_and_galaxy_growth} and~\ref{tracks}.}
\label{bimodality}
\end{figure*}

\begin{figure*} 
\noindent
\centerline{\hbox{
      \psfig{figure=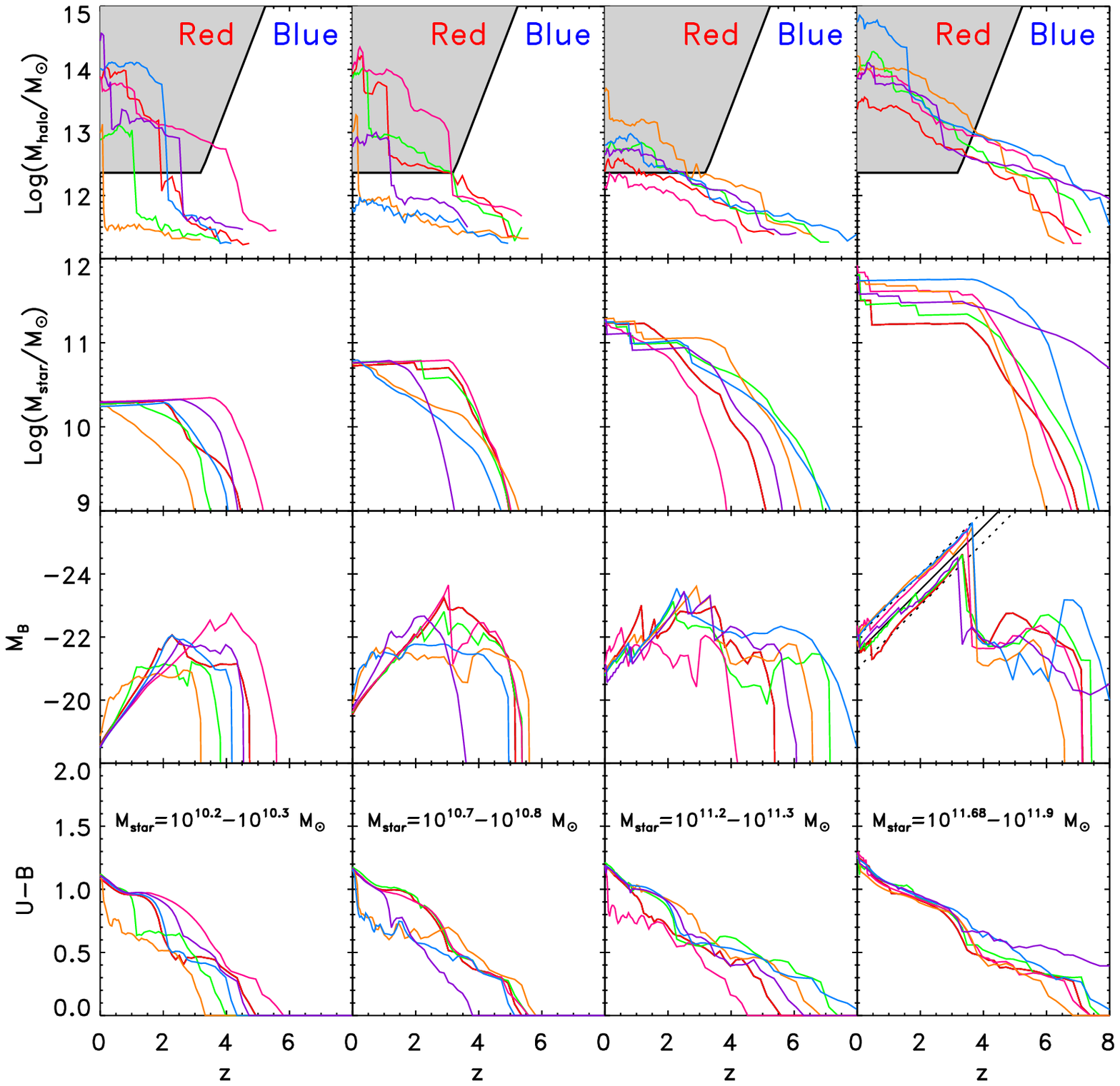,height=20.cm,angle=0}
  }}
\caption{Growth of halo mass (top) and stellar mass (bottom) for 24 galaxies 
randomly sampled from the four mass bins indicated (see Fig.~\ref{bimodality}). 
The growth curves refer to the most massive progenitors.
The bold black line in the upper panels is $M_{\rm crit}(z)$
above which cooling and star formation are shut down (Eq.~1). After $z\sim3$,
$M_{\rm crit}\sim M_{\rm shock} \sim 2\times 10^{12}M_\odot$.
After $M_{\rm halo}$ has crossed this threshold, the central galaxy ceases 
to form stars and passively turns red and dead.
Growth of the stellar mass after this point occurs only via mergers.
The sudden jumps in $M_{\rm halo}$ occur when a galaxy becomes a satellite in a larger halo.
The passive fading of giant ellipticals after $z\sim 3-4$ is well described by the simple relation
$M_B\sim -21.5-z\pm 0.5$ (solid line between the two dotted lines corresponding to $\pm 0.5\,$mag)}.
\label{halo_and_galaxy_growth}
\end{figure*}

\section{Shutdown and downsizing} 
\label{sec:entry} 

In the new model, the accretion of gas onto the central galaxy is shut down
once $M_{\rm halo}> M_{\rm crit}$.
Both in the standard and the new model, as common in other SAMs, 
the accretion of gas onto satellite galaxies is not allowed.\footnote{Admittedly, this procedure
may be too strict and should be revised, especially in small haloes with no 
hot medium, where anyway it may be hard to distinguish between central 
galaxies and satellites (e.g. \citealp{cattaneo_etal07}; also see \citealp{mccarthy_etal07}).}
Figure~\ref{mgalvsmhalo} shows that the stellar
mass keeps pace with the halo mass up to $M_{\rm crit}$ and then breaks off
in the new model.
It also shows the effect of preventing the accretion of gas onto satellite galaxies, which are
shown by the point clouds in the lower right part of the diagrams.

In order to maximize the quenching effect, whenever the shutdown criteria
are satisfied we actually shut down star formation altogether, that is,
in addition to having the halo gas shock heated and to keeping it hot, we 
assume that the cold gas is removed from the galaxies and that it is added to the 
hot halo. This can be achieved in central galaxies by thermal evaporation
\citep{nipoti_binney07}, and in satellite galaxies by ram-pressure stripping 
\citep[e.g.][]{bahcall77,gallagher78}. 
\citet{cattaneo_etal06} demonstrated that this robust shutdown leads to 
a good fit to the observed colour-magnitude distribution 
(Fig.~\ref{colour_histograms}, centre). In the absence of such an explicit
shutdown the standard model predicts too many luminous and blue galaxies
(Fig.~\ref{colour_histograms}, left).

The only slight disagreement between the predictions of the simple model of 
shutdown by halo mass
(red curves in the panel marked `no additional bulge mass criterion')
and the SDSS data (black smoothed histograms) is a small excess of blue 
galaxies at $-22.25\lsim M_r$, which arises from the absence
of quenched central galaxies with $M_{\rm star}\lsim 1-2\times 10^{11}$, 
since their haloes are below the shock-heating scale.
This is the range where the elliptical 
population is dominated by rather discy configurations, which do not show X-ray 
emission in excess of the contribution of discrete sources \citep{bender_etal89}.
These galaxies could have been possibly quenched by another mechanism,
such as quasar feedback after gas-rich mergers
\citep{springel_etal05,hopkins_etal07}, following the notion that discy 
ellipticals emerge from wet mergers (e.g. \citealp{cox_etal06}), which also
trigger quasar activity
(\citealp{toomre_toomre72,cattaneo_etal99,kauffmann_haehnelt00}; 
\citealp{cattaneo_etal05b}a,b; \citealp{springel_etal05b,hopkins_etal05}). 
We do not attempt here to model quasar feedback in detail.
We simply mimic its effect by stopping the accretion onto a galaxy
whenever it is dominated by its bulge component, even if it is still in a halo 
with $M_{\rm halo}<M_{\rm crit}$. This can occur only after a major merger or a 
sequence of minor mergers since the disc instability criterion does not permit the growth of bulges with  $M_{\rm bulge}>M_{\rm star}/2$
(see Section~2.3).
This additional quenching criterion slightly improves the fit 
to the SDSS colour-magnitude distribution near $M_r \simeq -21.5$
(the `new model' in Fig.\ref{colour_histograms}).
Note that the cold gas is not removed when a galaxy is quenched by this 
criterion alone.
While this is clearly a simplified {\it ad hoc} implementation of quasar
feedback, it has negligible effects at the bright end and at the faint end,
and it hardly affects the main results of our current analysis.

One could conceive an alternative scenario where quasar feedback after 
galaxy mergers is the main shutdown mechanism \citep[e.g.][]{hopkins_etal07c},
although it would have a hard time explaining the long-term maintenance of quenching. 
There is some observational evidence for such a mechanism in action
\citep{schawinski_etal07}. Modelling this specific scenario is beyond the scope of the present paper, 
where we focus on the effect of halo quenching on downsizing.

In summary, the new model stops the accretion of gas onto a galaxy 
when at least one of the following three conditions is satisfied:
(a) $M_{\rm halo}>M_{\rm crit}$, given by Eq.~1, or 
(b) $M_{\rm bulge}>M_{\rm star}/2$,
(c) the galaxy has become a satellite.
When the quenching is by condition (a), the cold gas is explicitly removed,
but even in the other cases the starved galaxies rapidly exhaust their gas
and stop making stars.
Fig.~\ref{bimodality} illustrates via different colours the role of 
these three different mechanisms in producing the galaxy bimodality. 
We see that  halo-mass quenching (a) is the main mechanism for $M_{\rm star}\gsim 2\times 10^{11}M_\odot$, and that 
the bulge quenching criterion (b) plays a role
near $M_{\rm star}\sim 10^{11}M_\odot$,
while most low mass galaxies are both quenched by halo mass (a) and satellites (c).
Only 3\% of the red sequence at $M_{\rm star}>10^{10}M_\odot$ is made of galaxies that are quenched only because they are satellites (c).
Although 57\% of the red sequence is composed of galaxies with $M_{\rm bulge}>M_{\rm star}/2$, 
only 24\% of red galaxies are quenched by criterion (b) only; 
$\sim 10$\% of the galaxies quenched by  criterion (b) only are red because they are dusty.
These galaxies are remnants of recent gas-rich mergers. They have ceased to accrete gas, but they have not yet ceased to make stars.
They account for $\sim 2$\% of the red sequence at $M_{\rm star}>10^{10}M_\odot$.
About 8\% of the red sequence at $M_{\rm star}>10^{10}M_\odot$ is made of galaxies that have not been explicitly quenched by any of 
the three mechanisms above (blue points in Fig.~\ref{bimodality}). The breakdown of this 8\% is as follows.
About 7\% are spiral galaxies that have naturally run out of gas without
an obvious external trigger, and the other 1\% are red because they are dusty.
In conclusion, 3\% of the red sequence, a rather low fraction, is red because of dust.

The shutdown of star formation when the halo mass grows above 
$M_{\rm crit}$ introduces the notion of an `entry mass' to the red sequence,
$M_{\rm star}^{\rm crit}$, that is the typical stellar mass of the central galaxy 
of an $M_{\rm crit}$ halo \citep{faber_etal07}. 
The entry mass is predicted to be constant after $z\sim 2$ and higher at higher redshifts \citep{dekel_birnboim06}.
Its value at $z\lsim 2$ is under 10\% of $M_{\rm shock} \sim 10^{12}M_\odot$, 
given the universal baryonic fraction of $\sim 0.15$ and
assuming that at least half of the gas is prevented from making stars
due to stellar feedback.

\begin{figure*} 
\noindent
\centerline{\hbox{
      \psfig{figure=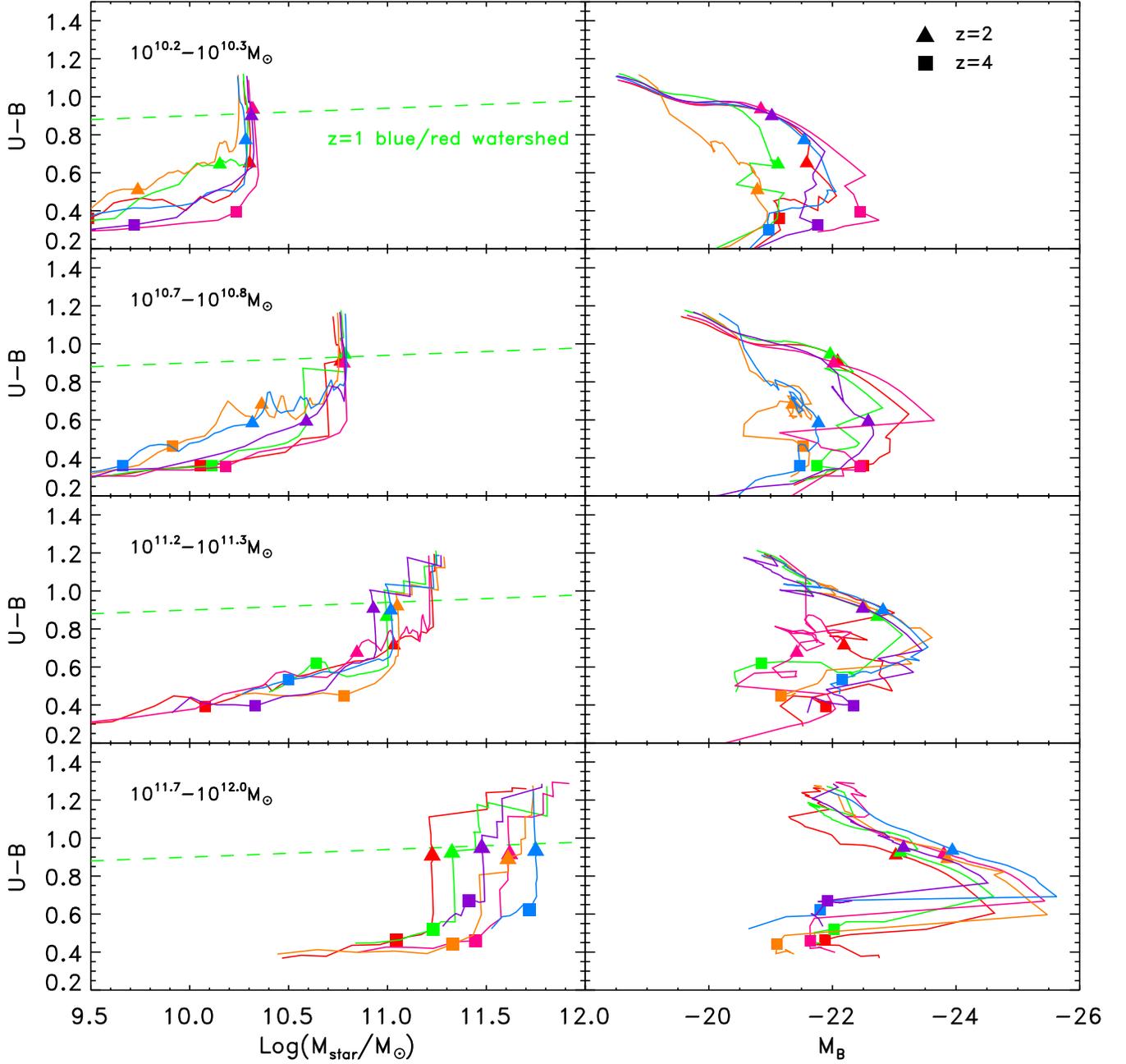,height=20.cm,angle=0}
  }}
\caption{Time evolution in the colour-mass diagram for galaxies that are
currently in four different mass bins along the red sequence, which are the
same as in Fig.~\ref{bimodality} and Fig.~\ref{halo_and_galaxy_growth}. 
The history tracks refer to the most massive progenitors.
The colour assigned to each galaxy is the same as in 
Fig.~\ref{halo_and_galaxy_growth}.
The green dashed line marks the green valley that separates
the red sequence  from the blue cloud at $z\sim 1$ in our model (see 
Fig.~\ref{bimodality_emergence}).
The symbols refer to two different redshifts, as indicated, and show the pace at which 
galaxies move along the colour-mass tracks.
Ignoring bursts of star formation, galaxies get steadily more massive and 
redder until they turn much redder in a short time once they are quenched. 
Small satellites are quenched and then keep a fixed mass. 
Massive galaxies continue to grow by dry mergers after they are quenched.
The massive galaxies with $M_{\rm star}\gsim 10^{11.7}M_\odot$ 
have already been quenched by $z=3$. 
Dry merging is not apparent at $M_{\rm star}\lsim 10^{11}M_\odot$.
It becomes an important growth mechanism above this mass.
We have plotted this figure assuming that all galactic discs are 
observed face-on.
}
\label{tracks}
\end{figure*}

We illustrate the connection between shutdown, entry mass and downsizing 
by comparing the halo and stellar mass growth histories of galaxies with
different final stellar masses (Fig.~\ref{halo_and_galaxy_growth}).
We consider four mass bins along the red sequence in the colour-mass 
diagram (corresponding to the four red parallelograms in Fig.~\ref{bimodality})
and select six galaxies at random from each bin. The mass intervals corresponding
to the four bins cover a factor of forty in stellar mass,
from $\sim 0.2M_{\rm star}^{\rm crit}$ to $\sim 8M_{\rm star}^{\rm crit}$. 

All galaxies display an initial phase, in which the stellar mass grows rapidly,
followed by shutdown and a passive phase, in which the stellar mass remains 
essentially constant (Fig.~\ref{halo_and_galaxy_growth}). 
The transition between the two phases is accompanied by a very fast reddening 
of the $U-B$ galaxy colour. This reddening is recognizable as a vertical jump upward
in a diagram of colour vs. stellar mass (Fig.~\ref{tracks}, left column).
It can also be seen in the galaxies' evolutionary colour-magnitude tracks (Fig.~\ref{tracks}, right
 column),
as it corresponds to the point where the galaxies start fading and reddening.
In 23 out of 24 galaxies,
the transition from blue to red coincides
with the time when the halo mass crosses $M_{\rm crit}$. 
A slight decrease of the stellar mass in the passive phase is due to 
the death of old stars not being replaced by new ones.
Growth in this phase is by dry mergers only, seen 
as sudden upwards turns after $M_{\rm star}(z)$ has flattened 
(Fig.~\ref{halo_and_galaxy_growth}), in contrast with 
the smoother growth of $M_{\rm star}$ due to star formation.

The most massive galaxies have final stellar masses 
in the range $10^{11.7}M_\odot\lsim M_{\rm star}\lsim10^{11.9}M_\odot$.
In these galaxies, $M_{\rm star}\gg M_{\rm star}^{\rm crit}$ because 
their haloes started forming very early and crossed $M_{\rm crit}$ 
when the entry mass was larger than 
$M_{\rm star}^{\rm crit}\sim 10^{11}M_\odot$.
The upturn of $M_{\rm crit}(z)$ at $z \geq 3$ causes most of these giant galaxies 
to become red-and-dead almost simultaneously at $z\simeq 3.5-4$.
Their post-quenching luminosity evolution is reasonably well described by a simple relation of the form
\begin{equation}
M_B\sim -21.5-z\pm 0.5
\end{equation} 
(diagonal solid line and $\pm 0.5\,$mag dotted lines in Fig.~\ref{halo_and_galaxy_growth}).
The haloes of these galaxies correspond to the sites of galaxy clusters at low redshifts.

Frequent mergers inside massive haloes 
provide an additional mechanism for the growth of giant galaxies that 
can raise their mass up to $M_{\rm star}\sim 10^{12}M_\odot$.
Indeed, five out of our six massive galaxies 
grow by a factor of $\sim 2$-$3$ through dry mergers after entering
the red sequence.
This late merging agrees with semianalytic modelling calculations by 
\citet{delucia_etal06} and \citet{delucia_blaizot07}.
A note of caution regarding the comparison of these predictions to observations
is that the light associated with the growing stellar mass at the centres of galaxy clusters
may be spread out across a non-negligible fraction of the cluster volume. 
Combined with observational limitations, this may lead to
an underestimate of the stellar mass in brightest cluster galaxies at the
centers of clusters (\citealp{gonzalez_etal05,lauer_etal07,faber_etal07}).

The upper-intermediate mass bin in Fig.~\ref{halo_and_galaxy_growth}
($10^{11.2}M_\odot<M_{\rm star}<10^{11.3}M_\odot$) 
is populated by central galaxies 
in which $M_{\rm halo}$ has crossed $M_{\rm crit}$ at $z\lsim 3$,
where $M_{\rm crit}=M_{\rm shock}$ (Eq.~1). 
All of these galaxies enter the red sequence with a same stellar mass of
$M_{\rm star}^{\rm crit}\sim 10^{11}M_\odot$.
In both the high and the upper-intermediate mass bin there is a strong 
correlation between the final stellar mass and the redshift at 
which the galaxies become red.
Galaxies that enter the red sequence earlier end up being more massive at $z\sim 0$, 
even when the entry mass is similar, because they have more time to grow 
through dry mergers along the red sequence.

In contrast, the two lower mass bins 
($10^{10.7}M_\odot<M_{\rm star}<10^{10.8}M_\odot$
and $10^{10.2}M_\odot<M_{\rm star}<10^{10.3}M_\odot$)
are dominated by satellite galaxies, identified in
Fig.~\ref{halo_and_galaxy_growth} by an abrupt big
rise in their halo mass growth as they fall into a massive host halo.
As satellites, they have a lower merging rate, with the growth by dry mergers
along the red sequence playing a role only in a couple of the low-intermediate
mass galaxies, and in none of the low-mass galaxies.
For most of the low-mass galaxies, the transition to the red sequence
corresponds to a merger of their dark matter halo into a halo more massive
than $M_{\rm crit}$, causing them to shut down as satellites.
As the timings of these events are not strongly correlated with the halo mass 
before merging, the low-mass galaxies display a wide spread in quenching times, which are not directly related to final mass.

One lower-intermediate mass galaxy (light blue) resides in a halo less 
massive than $M_{\rm crit}$ (and thus appears as one of the 7\% blue symbols
in Fig.~\ref{bimodality}). This galaxy has not been quenched by any of 
the three explicit mechanisms, but rather ran out of gas by continuous 
star formation.

One of the low-mass galaxies (fuchsia) has been the central galaxy of a halo that reached nearly 
$10^{13}M_\odot$ before it crossed $M_{\rm crit}$ at $z\gsim 3$.  
However, this galaxy remained small because it was quenched before it had time to convert into stars the gas that had accreted and was accreting onto it.
 
To conclude, 
Fig.~\ref{halo_and_galaxy_growth} demonstrates that shutdown at a 
constant critical halo mass introduces a rather constant
entry mass for {\it central} galaxies into the red sequence.
Central galaxies that enter the red sequence earlier end up more massive
because they have more time for subsequent growth by dry merging.
Note that the downsizing of $M_{\rm crit}$ at high redshift predicted by
\citet{dekel_birnboim06} in itself induces a downsizing in the entry mass,
which enhances the effect of archaeological downsizing in today's giant
ellipticals.
Dry merging is important in the growth of massive red galaxies with
$M_{\rm star} \geq 10^{11}M_\odot$.
Galaxies that become satellites enter the red sequence with masses lower than the characteristic entry mass,
and they dominate today's population at $M_{\rm star}\ll 10^{11}M_\odot$. 

\begin{figure*}
\noindent
\centerline{\hbox{
      \psfig{figure=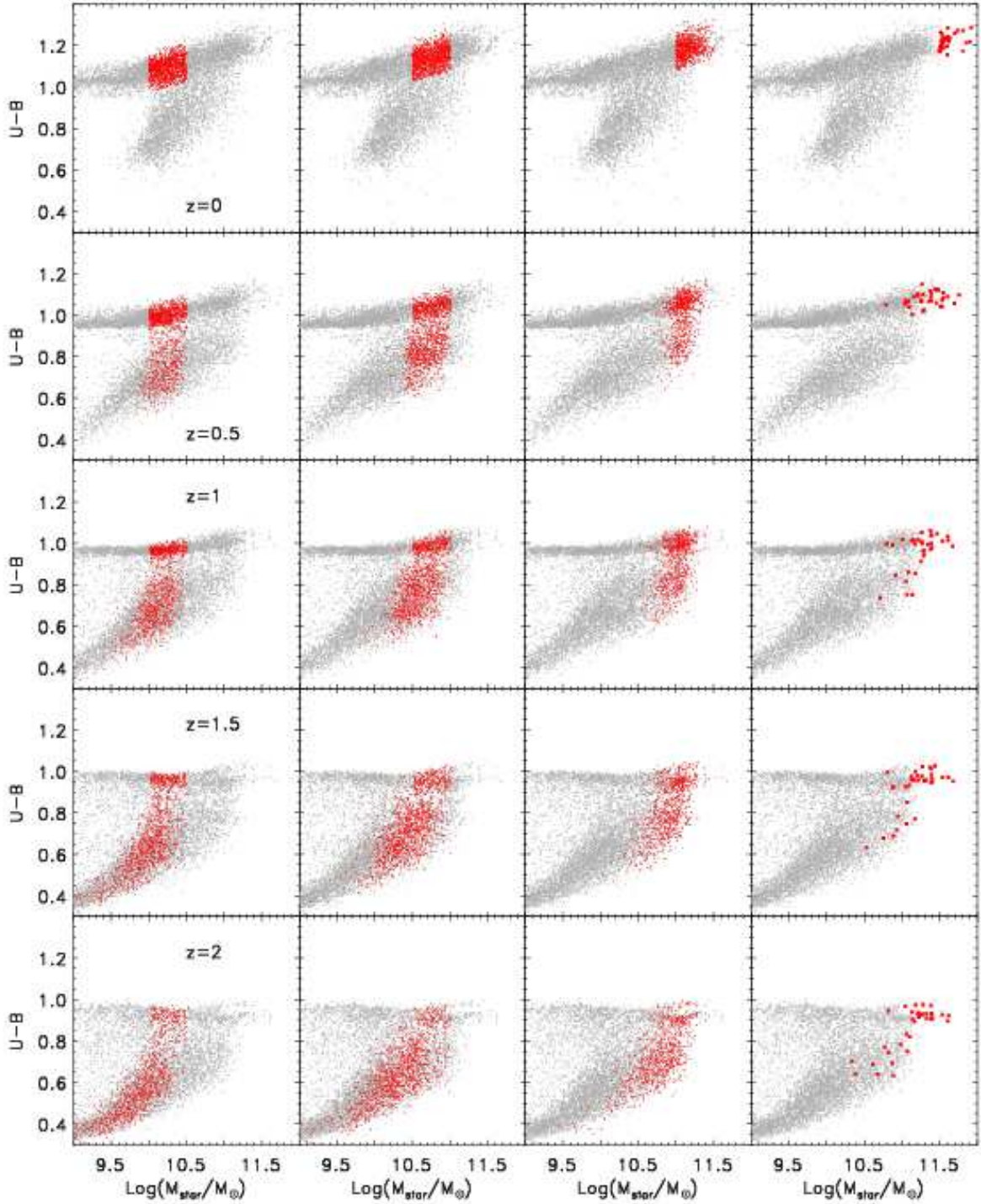,height=20.cm,angle=0}
  }}
\caption{Evolution of  the colour-mass diagram from $z\sim 2$ to $z\sim 0$. 
The grey dots represent 8,000 randomly selected galaxies 
at $z=0$ (and all the giant galaxies), 
and their main progenitors at higher redshifts.
The galaxies with $M_{\rm star}>10^{10}M_\odot$ on the red sequence at $z=0$
are divided into four bins according to their final stellar mass,
as marked by the red symbols in the four columns of the upper row.
The red symbols in the lower panels highlight the corresponding main
progenitors at the different redshifts.}
\label{bimodality_emergence}
\end{figure*}

\section{Downsizing in the assembly of the red sequence}
\label{sec:assembly} 
 
In this section, we show that 
the shutdown  above a critical halo mass produces downsizing in the buildup of the red sequence,
in the sense that the high-mass end is populated at earlier times.
For this purpose,
we consider in Fig.~\ref{bimodality_emergence} a large randomly selected sample 
of 8,000 galaxies, except that {\it all} the giant galaxies with 
$M_{\rm star}> 10^{11.5}M_\odot$ are included for better statistics.
We plot a series of colour-mass diagrams, which show these galaxies at $z=0$,
and their main progenitors at higher redshifts.
The red symbols highlight the galaxies that end up on the red sequence
in four distinct mass intervals, ${\rm Log}M_{\rm star}=[10,10.5],\,[10.5,11],\,[11,11.5],\,[11.5,12]$.

The overall population exhibits the same behaviour seen in
Fig.~\ref{halo_and_galaxy_growth}. The distinction between a blue cloud and a 
red sequence is already present at $z\gsim 2$ 
(in agreement with FIRES observations; \citealp{giallongo_etal05}), 
with about 60\% of the most massive galaxies
already on the red sequence at that redshift.
Upper intermediate mass galaxies grow along the blue cloud
until they reach the entry mass, 
near the upper mass limit for the blue cloud,
where they turn red.
The most massive galaxies climb up to the top of the blue sequence, turn red,
then continue their growth by dry merging along the red sequence.

 \begin{figure*}
\noindent
\begin{minipage}{8.4cm}
  \centerline{\hbox{
      \psfig{figure=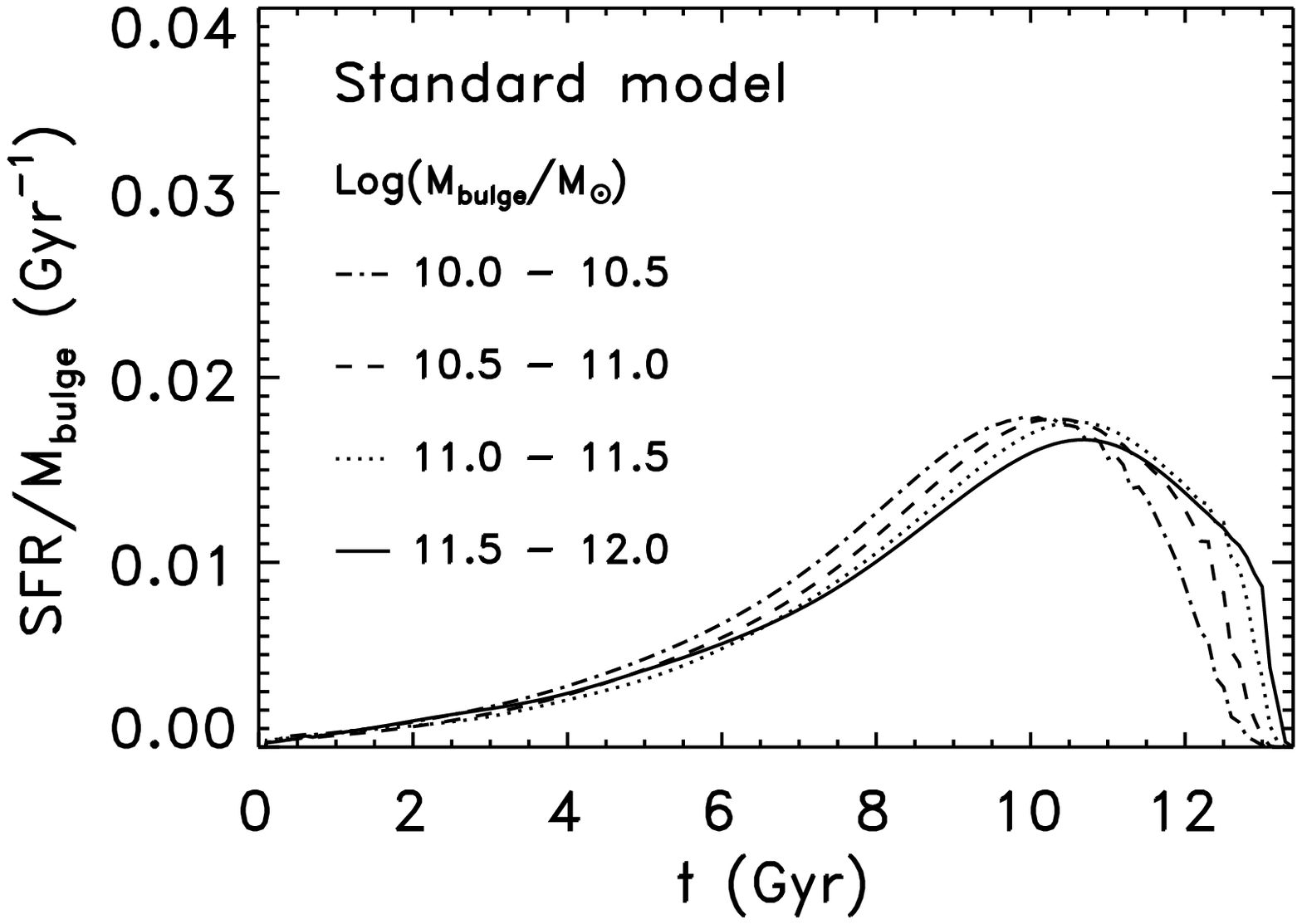,height=6.cm,angle=0}
  }}
\end{minipage}\    \
\begin{minipage}{8.4cm}
  \centerline{\hbox{
      \psfig{figure=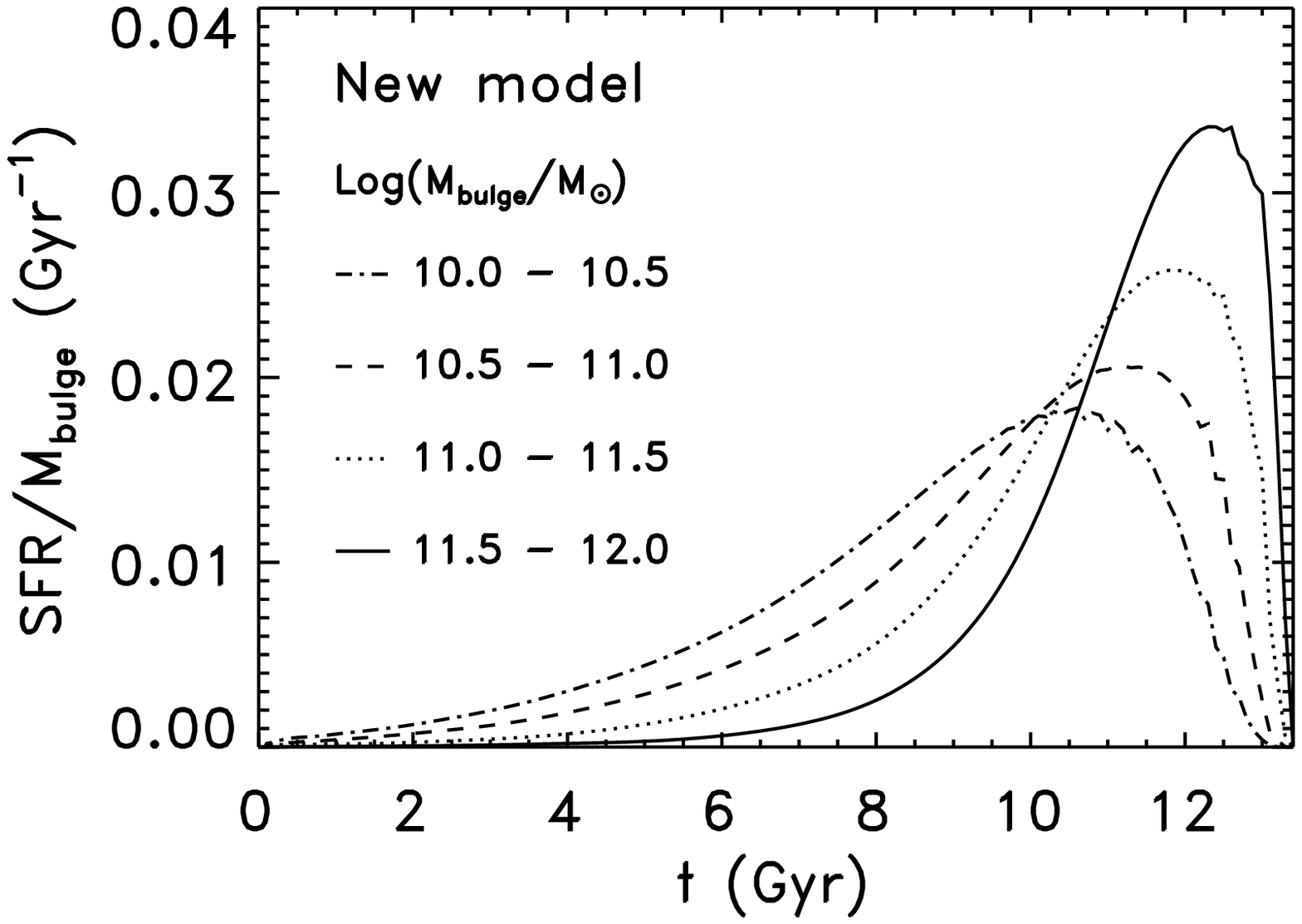,height=6.cm,angle=0}
  }}
\end{minipage}\    \
\caption{The distribution of stellar ages as a function of mass for
bulge-dominated galaxies,
obtained by averaging the specific SFR histories in the given mass bins. 
The specific SFR is inferred from the age distribution of all the stars in today's
galaxies (not just those in their main progenitors). Left: the standard model. Right: the new model, in which star 
formation is shut down above a critical mass and in bulge-dominated central galaxies.}
\label{thomas}
\end{figure*}

The mass growth due to dry mergers is a factor of $\sim 2.5$ in the highest 
mass bin and $\sim 1.7$ in the upper-intermediate mass bin. 
It becomes negligible below $10^{11}M_\odot$.
This transition from wet to dry mergers may be related to the observed
division of elliptical galaxies into two major types, giant boxy 
non-rotating ellipticals versus the smaller discy rotating ellipticals.
Indeed, the origin of this distinction has been proposed to be the
difference between dry merging along the red sequence and recent wet mergers
along the blue cloud
\citep{bender_etal88,bender_etal89,nieto_bender89,nieto_etal91,
bender_etal92,kormendy_bender96,tremblay_merritt96,gebhardt_etal96}.  
The transition from wet to dry mergers may also be related to the division 
of ellipticals according to their inner density profiles, where giant 
ellipticals show cores and less massive ones show power-law cusps, 
possibly reflecting gas-rich mergers \citep{faber_etal97,lauer_etal07}. 
This transition is observed at $M_V\simeq-20.5$, which is 
$\sim 10^{10.8}M_\odot$ 
(assuming $M/L_V\simeq 6\times 10^{-0.092(M_V+22)}M_\odot/L_\odot$
based on the $M/L$ estimates of \citet{gebhardt_etal03}).  
This is indeed close to our estimated entry mass of 
$M_{\rm star}^{\rm crit}\sim 10^{11}M_\odot$, below which the growth is mainly
by wet mergers.

The same transition mass is typically the minimum entry mass for central
galaxies. It is also comparable to the bright edge of the blue cloud.
The fact that this bright edge is constant with time reflects the fact that
the entry mass for central galaxies is roughly constant after $z \sim 2$.

The low-mass bin and the lower intermediate mass bins, which are dominated by satellites,
are mainly populated after $z \sim 1$. 
Low mass galaxies make an abrupt transition from blue to red
at masses lower than the minimum entry mass for central galaxies.
This is consistent with the notion that they are quenched when they
fall into larger haloes and become members of groups or clusters.
 
We notice that at $z\sim 2$ the red sequence has a rather constant colour
independent of mass, and that a tilt develops gradually in time.
The origin of this effect, and its relation to downsizing,
will be addressed elsewhere.

\section{Archaeological downsizing in stellar ages}
\label{sec:thomas}  

Perhaps the most useful way to measure downsizing is by comparing the
distribution of stellar ages in galaxies of different masses.
\citet{thomas_etal05} deduced stellar ages from the absorption line indices
H$\beta$, Mg $b$ and $<$Fe$>$ for a sample of 124 early-type red sequence galaxies.
They fitted the distribution of stellar ages in galaxies of a given stellar
mass using a Gaussian, with the mean time and standard
deviation as free parameters. Their main finding, as read from their 
Fig.~10, is that the mean stellar age is increasing with mass,
while the scatter is decreasing with mass.
They also find that the mean stellar age is increasing with environment density
at a given stellar mass.
It is important to realize that the measured age spreads 
are inferred from [$\alpha$/Fe] and not from detailed modelling of galaxy spectra.
Moreover, the spread in the times at which different galaxies attain their peak
SFR may be significantly larger than the age spread in each galaxy, 
which is what [$\alpha$/Fe] measures.
Taking these differences into account is important when we compare the 
age spread estimated by \citet{thomas_etal05}
with those obtained from the mean star formation histories in our model.

To compare our model with the results by \citet{thomas_etal05}, 
we should consider {\it all} the stars that end up in a given red galaxy, and not 
just those that formed in its most massive progenitor.
Fig.~\ref{thomas} plots the predictions of our model for the distribution of 
stellar ages in the spheroids of bulge-dominated galaxies.
We compare the results for the standard model (no shutdown, except in
satellites) and the new model 
(shutdown of star formation in satellites, in massive haloes and in 
bulge-dominated galaxies). 
Once again, we split the red galaxies into four stellar mass bins. 
We see that, in the standard model, galaxies of all masses have similar
star formation histories.
The weak downsizing signature probably reflects the natural downsizing 
of dark haloes in which star formation occurs only above a minimum halo mass
\citep{neistein_etal06}.
In contrast, the new model shows a strong downsizing effect, where more
massive galaxies form stars earlier and over a shorter time span, as observed.

In both models, the SFR histories for all masses are predicted to have long 
tails extending to the present time.
This has also been seen in the semi-analytic simulations by 
\citet{delucia_etal06}.
Note, however, that the curves in Fig.~\ref{thomas} are averages over 
many galaxies. Individual galaxies have a narrower age spread. 
The general skewness of the SFR toward early times reflects the overall 
trend of the star formation history in the Universe, where roughly half the
stars are made before a look-back time of $t=8\,$Gyr ($z=1$). 
In the new model, the SFR in massive galaxies is truncated by crossing  
$M_{\rm crit}$ early. 
Less massive central galaxies pass through $M_{\rm crit}$ later, so their
star formation stops later.

The two lowest mass bins have the widest age spreads because satellites fall
into bigger haloes and get quenched over a wide range of times. 
Small differences between the two lowest mass bins reflect both the residual 
effect of central galaxies in the lower intermediate mass bin and the 
tendency of more massive satellites to fall into massive haloes earlier.

\begin{figure}
\noindent
\centerline{\hbox{
      \psfig{figure=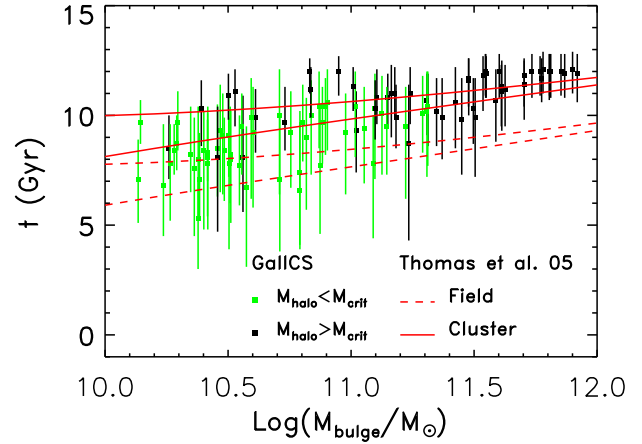,height=6.4cm,angle=0}
  }}
\caption{The median stellar age of bulges in bulge-dominated (E/S0) galaxies
as a function of mass.
The symbols and error bars correspond to the look-back time by which 
$(50\pm34)\%$ of the bulge stellar mass has formed. To mimic environment density,
the green and black symbols refer to galaxies in haloes below and above 
the critical mass $\sim 2\times 10^{12}M_\odot$, respectively.
The red lines show in comparison the $\pm \sigma$ age spread as  
inferred by \citet{thomas_etal05} for E/S0 galaxies in field and cluster 
environments.}
\label{agevsmass}
\end{figure}

Fig.~\ref{agevsmass} compares the results of the new model with the mean 
stellar ages and internal age spreads inferred by \citet{thomas_etal05},
including the environment dependence. 
The model largely reproduces the observed effects.
In the two environments, the model mean stellar ages increase with galaxy mass
in general agreement with the observed trends, except that the model has a 
slight tendency to overestimate the stellar ages in the field (see below),
but the big difference between environments may be an artifact of the
small sample used (D. Thomas, private communication).
In our model, the stellar populations of the most massive ellipticals are as 
old or even slightly older than those in
\citet{thomas_etal05}, even in the densest environments. 
The model has no trouble making old-enough stars in massive galaxies.
At any given galaxy mass, the age of the stellar population is higher in a cluster environment than in the field.

The internal spread in stellar ages is narrower for more massive galaxies,
both in the model and the data.  
While in the model the most massive galaxies formed all their stars in a short 
time span $\lsim 2\,$Gyr, this time span is not as brief as that inferred by 
\citet{thomas_etal05} from their measured values of 
$[\alpha/{\rm Fe}]$, which is $\sim 0.3\,$Gyr.
At a time of $t \sim 2\,$Gyr after the Big Bang, the characteristic dynamical 
time for flowing at the virial velocity from the virial radius to the halo
centre is $\sim 0.4\,$Gyr. The observed duration thus implies a
massive burst of star formation on the shortest possible dynamical time
while the model predicts a longer duration, which is comparable to the Hubble time 
at that epoch.
A burst as short as inferred by \citet{thomas_etal05} requires a process that accelerates
the SFR and then rapidly shuts it down, as it could possibly be obtained by 
major mergers, quasar feedback, or shocked accretion \citep{birnboim_etal07}.
Such a process is not incorporated in our current model.

The observed mean stellar ages are expected to be slightly lower than 
the model predictions for several reasons. First, the observed
light-weighted average ages are younger than the model mass-weighted averages
because the younger stars are brighter and dominate the light. 
Furthermore, younger populations have stronger Balmer equivalent widths, 
which dominate the total equivalent widths, and thus add to the 
brightness effect just mentioned, making light-weighted ages still younger. 
Finally, young tails in the age distribution, if present in real galaxies,
would pull the peak of the Gaussian fit toward younger ages. 
Considering these limitations, the agreement between model and data is good.


\section{Downsizing in star formation vs. upsizing in assembly}
\label{sec:upsizing} 

\begin{figure}
\noindent
\centerline{\hbox{
      \psfig{figure=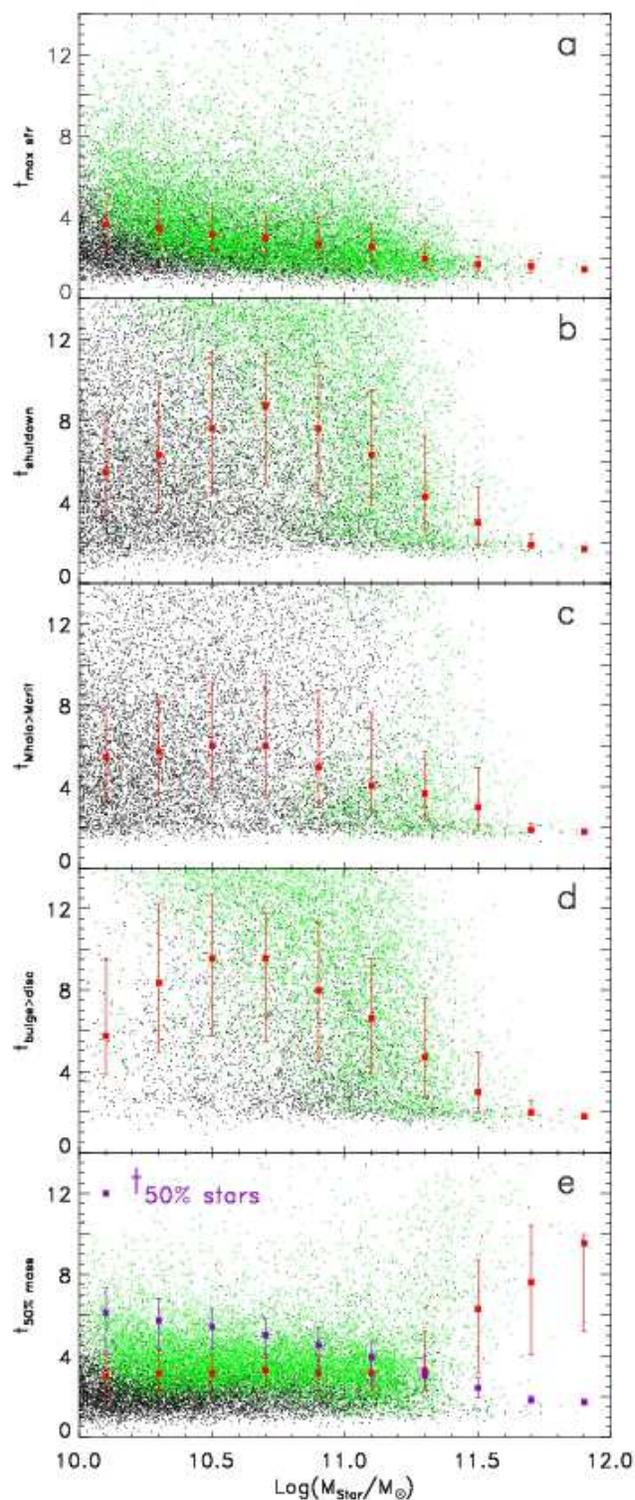,height=21.cm,angle=0}
  }}
\caption{Times characterizing the SFR and its shutdown versus final stellar
mass.
(a) The time of peak SFR $t_{\rm max\,sfr}$. 
(b) The time of star formation shutdown $t_{\rm shutdown}$.
(c) The time at which $M_{\rm halo}$ crosses $M_{\rm crit}$, 
$t_{\rm Mhalo>Mcrit}$. 
(d) The time at which the bulge mass becomes larger than the disc mass 
$t_{\rm bulge>disc}$.  
(e) The assembly time at which the main progenitor attains $50\%$ of the final 
stellar mass, $t_{\rm 50\%}$ (red symbols). Also shown is the time at which
half of the stars have formed (purple symbols).
The times, measured from the Big Bang, are evaluated by tracking the 
main progenitor of each galaxy back in time.
Central galaxies today are marked green and satellites are marked black.
Red symbols with error bars are median values and lower and upper quartiles 
in ten mass bins.
} 
\label{times}
\end{figure}

In Fig.~\ref{times} we extend the analysis of the new model
to other times that characterize the SFR and its shutdown. 
These are the times 
(a) of peak SFR, 
(b) of shutdown,
(c) at which $M_{\rm halo}$ crosses $M_{\rm crit}$,
(d) at which the galaxy becomes bulge dominated, and
(e) at which the main progenitor attains half the final galaxy stellar mass.
Each characteristic time can be interpreted as showing 
either downsizing or upsizing depending on whether it is 
monotonically decreasing or increasing as a function of 
final stellar mass.  
We shall see that all the above time indicators referring to star formation
show a general downsizing trend, as opposed to the upsizing trend that 
the model predicts for the analogous time indicators that characterize 
mass assembly.

It is worth noting the differences between Fig.~\ref{times} and 
Fig.~\ref{agevsmass}.
First, the times $t$ in Fig.~\ref{times} are from the Big Bang, while 
in Fig.~\ref{agevsmass} they were lookback times, that is $t_0-t$, where $t_0=13.7\,$Gyr is the present age of the Universe
in the cosmology used for this study.
Second, Fig.~\ref{agevsmass} is constructed from a small number of  
galaxies, where the error bars correspond to the internal stellar age 
spreads in the individual galaxies, while Fig.~\ref{times} shows the 
entire red galaxy population in the new model.
Third, the galaxies in Fig.~\ref{agevsmass} are all red bulges of
bulge-dominated galaxies, while
Fig.~\ref{times} shows all the red galaxies, and the times contain the 
contribution of the disc component if present.
Fourth, Fig.~\ref{agevsmass} refers to all the stars in today's bulges, while
Fig.~\ref{times} refers only to the stars in the galaxies' main progenitor back in time
(the purple symbols in Fig.~\ref{times} are an exception).
 
\begin{figure}
\noindent
\centerline{\hbox{
      \psfig{figure=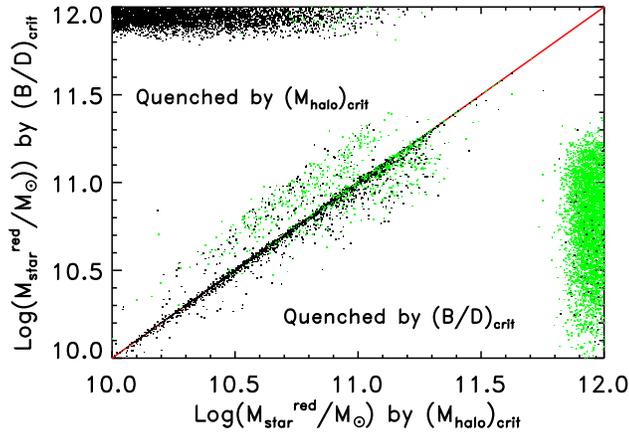,height=6.4cm,angle=0}
  }}
\caption{Role of the two quenching criteria.
Each point shows the stellar masses of a galaxy when it became 
bulge dominated ($y$-axis) and when its halo crossed $M_{\rm crit}$ ($x$-axis). 
Galaxies that crossed $M_{\rm crit}$ before (after) they became bulge-dominated 
lie above (below) the diagonal red line.
Green and black points refer to central and satellite galaxies at $z=0$, 
respectively.
Galaxies that never become bulge-dominated, mainly satellites, are put
near the top edge of the plot. 
Galaxies that never cross the critical halo mass, mainly centrals, are put near
the right edge of the plot. 
}
\label{bulgevsmcrit}
\end{figure}

The time of maximum SFR, $t_{\rm max\,sfr}$, is shown in Fig.~\ref{times}a.
In most red galaxies, it occurred within the first $\sim 4\,$Gyr from the Big Bang.
There is a clear but weak downsizing trend in the epoch of maximum SFR.
In the most massive galaxies, the SFR peaked at $t\sim 1-2\,$Gyr, while at 
low masses there is a larger spread in $t_{\rm max\,sfr}$.
We thus recover downsizing features similar to the `archaeological' downsizing
obtained from the stellar age distribution in red galaxies at $z\sim 0$.

For a fixed final stellar mass, the SFR peak occurs earlier in satellite 
galaxies than in central galaxies. 
This is because low-mass central galaxies 
are in low mass haloes in which star formation does not start early,
it is slowed down by supernova feedback, and it is not shut down early.
Satellites, on the other hand, much like high stellar mass galaxies,
can be part of massive haloes already at early times, so they can form 
stars early and also shut down early
(SDSS data, \citealp{blanton_etal06}; our new model, \citealp{cattaneo_etal06}).

\begin{figure}
\noindent
\centerline{\hbox{
      \psfig{figure=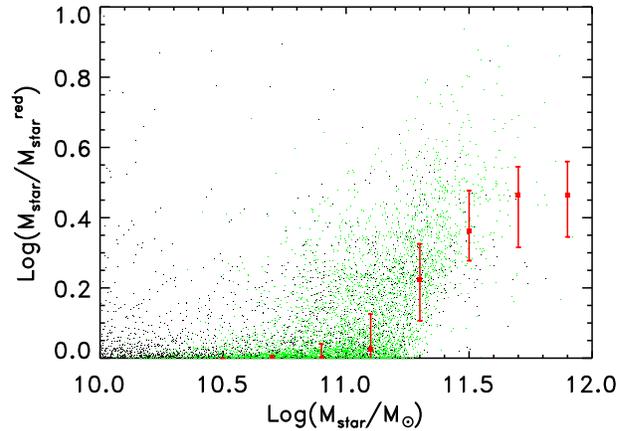,height=6.4cm,angle=0}
  }}
\caption{Growth by dry merging along the red sequence versus stellar mass. 
It is measured by the ratio between the final stellar mass of a red galaxy 
and the mass it had when it was quenched by one of the three shutdown 
mechanisms of our model.
Green and black points refer to central and satellite galaxies at $z=0$, 
respectively. The red symbols with error bars show the median values of
$M_{\rm star}/M_{\rm star}^{\rm red}$ and the lower and upper quartiles 
in logarithmic bin of $0.2\,$dex in $M_{\rm star}$.
}
\label{redgrowth}
\end{figure}

The time of shutdown in the most massive progenitor, $t_{\rm shutdown}$, is
shown in Fig.~\ref{times}b.
In $\sim 7\%$ of the red galaxies with $M_{\rm star}>10^{10}M_\odot$ 
the formation of stars has never been quenched by any of the 
explicit criteria applied in the new model.
These galaxies have turned red simply because they have exhausted their 
gas reservoir. 
In addition, $\sim 2\%$ of the red sequence at $M_{\rm star}>10^{10}M_\odot$ 
is made of starbursts, which are red because they are dusty.
We have removed these unquenched and dusty galaxies from Fig.~\ref{times}b
and from the shown median. 
The general downsizing trend in Fig.~\ref{times}b above $\sim 10^{10.7} M_\odot$ 
is similar to Fig.~\ref{times}a, especially for the central galaxies, 
but the magnitude of the effect is stronger for $t_{\rm shutdown}$.
This is because intermediate mass galaxies not only reach their peak 
SFR somewhat later than more massive galaxies, but they also 
have tails of SFR extending to late times, $z\lsim 1$ (Fig.~\ref{thomas}).
The most massive galaxies undergo an intense burst of star formation when 
the Universe is only $\sim 1.5\,$Gyr old and then they are almost immediately 
shut down. In red galaxies of intermediate masses, the peak SFR is
at $t \sim 3\,$Gyr and the formation of stars drags on for $\sim 6\,$Gyr 
after the peak. 
The shutdown time does not show downsizing at low masses,
where the red population is dominated by satellites.

We recall that our model contains two quenching mechanisms. 
One is due to the shock heating in haloes once they become more massive
than $M_{\rm crit}$, and the other is a result of wet major mergers.
Mergers are assumed to simultaneously produce dominant stellar spheroids 
and trigger quasars at their centres, which then help removing the gas available 
for star formation in merger remnants (e.g., \citealp{springel_etal05}). 
The bulge-dominance criterion applied in the the new model is meant
to mimic quasar feedback of this sort.
We assess the relative importance of these two mechanisms in the new model
by splitting Fig.~\ref{times}b into two,
showing separately the times for crossing the critical halo mass 
$t_{\rm Mhalo>Mcrit}$ (Fig.~\ref{times}c) 
and for the bulge to become dominant $t_{\rm bulge>disc}$
(Fig.~\ref{times}d).
Galaxies that have never crossed the critical halo mass do not appear in 
Fig.~\ref{times}c, and galaxies that have never become bulge-dominated do 
not appear in Fig.~\ref{times}d.

Massive galaxies with $M_{\rm star}>10^{11}M_\odot$ 
and $t_{\rm shutdown}<3\,$Gyr 
in Fig.~\ref{times}b populate the same region of the diagram in 
Fig.~\ref{times}c and Fig.~\ref{times}d. These are 
central galaxies of massive haloes, which simultaneously obey 
$M_{\rm halo}>M_{\rm crit}$ and have a dominant spheroid due to intense merging.
Galaxies that are less massive tend to show up either in Fig.~\ref{times}c
or in Fig.~\ref{times}d, depending on whether they are satellites or
small central red galaxies.

The same three populations can be identified in Fig.~\ref{bulgevsmcrit},
which displays for every galaxy the stellar masses at the two characteristic
times: when it entered a halo more massive than the critical mass, and 
when it became bulge-dominated.
The cloud of black dots at the very top consists of 
satellite galaxies in massive haloes, which have never become bulge-dominated
because the merger rate is low for satellites.
This population corresponds to the black symbols in Fig.~\ref{times}b 
that survive in Fig.~\ref{times}c but not in Fig.~\ref{times}d.
The cloud of green dots at the very right consists of 
bulge-dominated central galaxies in haloes that are just below the 
critical mass. These galaxies correspond to the intermediate mass galaxies 
shown as green symbols in Fig.~\ref{bimodality}.
Without the bulge shutdown criterion, these galaxies would be on the 
blue cloud, and they would extend it to higher luminosities \citep{cattaneo_etal06}.
This population corresponds to the green symbols in Fig.~\ref{times}b 
that survive in Fig.~\ref{times}d but not in Fig.~\ref{times}c.
The third population comprises galaxies that both inhabit massive haloes {\it and} 
are bulge-dominated.  This population contains the most massive objects 
and is composed of both central and satellite galaxies. 
The distinction between central and satellite galaxies is not very useful here
because many satellites were central until recently.
A more useful distinction is between the galaxies that were first quenched
by the bulge criterion and the galaxies that were first quenched by the halo 
criterion (points below and above the diagonal respectively).
These galaxies obey the two quenching criteria almost back-to-back
because the two events are related to the hierarchical growth of dark matter 
haloes. 

The characteristic times considered so far are indicators of star-formation 
history.
Fig.~\ref{times}e displays instead an indicator of the assembly history, 
the time $t_{50\%{\rm\,mass}}$ at which $50\%$ of the final stellar mass has been 
assembled in a single objects (red symbols). 
The assembly time is roughly constant with mass for low and intermediate-mass
galaxies, and it actually reveals upsizing at the high-mass end.
This resembles the assembly upsizing of the main progenitor of dark matter
haloes in \citet{neistein_etal06}. 

In galaxies of low or intermediate mass, $t_{50\%{\rm\,mass}}$ is comparable
to $t_{\rm max\,sfr}$ of Fig.~\ref{times}a. 
The complete shutdown of star formation occurs when the Universe is 
$\gsim 8\,$Gyr old, but $\gsim 80\%$ of the final stellar mass is already 
in place in the most massive progenitor after the first $\sim 5\,$Gyr. 
Only little mass is added to galaxies of $M_{\rm star}<10^{11}M_\odot$ 
after this point, as the merger rate is not high in this mass range,
as demonstrated by Fig.~\ref{redgrowth}, 
which shows the mass 
growth factor after entering the red sequence as a function of the final 
stellar mass.
Moreover, accretion onto the haloes of small galaxies,
and therefore onto small galaxies themselves, is suppressed by tidal effects in the
vicinity of more massive haloes (Hahn, Dekel et al., in preparation).

In contrast to small galaxies, galaxies with $M_{\rm star}\gsim 10^{11}M_\odot$ 
experience substantial growth through dry mergers after they cease to form 
stars (Fig.~\ref{redgrowth}). Therefore, the most massive galaxies continue gathering mass for 
a much longer time. When the Universe is $\sim 7\,$Gyr old, they have 
barely put together half of their final stellar mass. 
 
The onset of upsizing in assembly is right above the minimum 
entry mass for central galaxies, near the mass separating the boxy and 
discy ellipticals (Section~4).
Our model hints to a simple explanation for this bimodality
of structural and kinematic properties of galaxies along the red sequence.
Discy ellipticals are the products of recent wet mergers of gaseous disks
along the blue cloud, in haloes of $M_{\rm halo}<M_{\rm crit}$. 
Boxy ellipticals, on the other hand, result from 
dry mergers along the red sequence, 
typically at the centres of haloes with $M_{\rm halo}>M_{\rm crit}$. 
The most massive ellipticals, which have only finished assembling $80\%$ 
of their final stellar mass in the last few billion years, are to be 
identified with the brightest cluster galaxies at the centres of Abell clusters.
Also see \citet{delucia_blaizot06} for a semi-analytic study.

Fig.~\ref{times}e also displays for comparison the time at which $50\%$ of 
the final stellar mass was formed (purple symbols), showing a 
star-formation downsizing similar to the trends seen in the other panels.
This figure demonstrates how upsizing in assembly coexists with downsizing
in star formation in the same hierarchical clustering scenario.

A second final remark on Fig.~\ref{times} concerns the difference between central and satellite galaxies.
Compared to satellites of the same mass,
in central galaxies the SFR peaks later, it shuts down later,
and the stellar mass assembles later (compare the green 
and black dots in Fig.~\ref{times}a, b and e).

\begin{figure}
\noindent
\centerline{\hbox{
      \psfig{figure=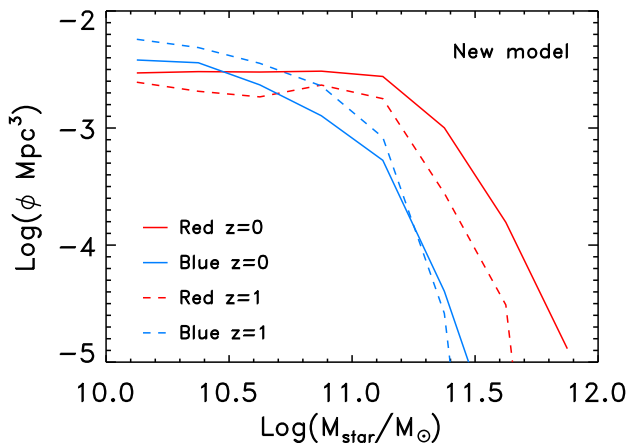,height=6.4cm,angle=0}
  }}
\caption{The evolution of the mass functions of red and blue galaxies from 
$z=1$ (dashed lines) to $z=0$ (solid lines) in our model with star 
formation shutdown.}
\label{massfunction_evolution}
\end{figure}

\section{Downsizing in the galaxy mass functions?}
\label{sec:mass_function} 

The time evolution of the mass functions of red and blue galaxies 
has been used as another indicator for downsizing in the formation of the red 
sequence.
The features considered, for example, are (a) the downward shift with time of 
the mass where the red and blue functions cross each other,
and (b) the indications for a late increase in the mass function of red galaxies at 
low masses compared to an early increase at high masses
(\citealp{bundy_etal05,borch_etal06,pannella_etal06}; also see the discussion
in \citealp{faber_etal07}).
In this section we highlight uncertainties in such arguments, both due to 
inherent difficulties in interpreting model mass functions and to the
large uncertainties that still exist in the data themselves.  
Our principal conclusion is that the mass functions of red and blue galaxies are poor indicators of
downsizing.

The observational uncertainties are discussed separately in the Appendix.
Here we concentrate on the predictions of the model and demonstrate how hard 
it is to use mass functions to infer downsizing or upsizing, even if the 
data were perfectly accurate.  

The evolution of the mass functions of red and blue galaxies in the new
model from $z=1$ to $z=0$ is shown in Fig.~\ref{massfunction_evolution}.  
The red mass function has grown at all masses, while the
blue function has decreased or remained constant.
Below $M_{\rm star}^{\rm crit}$ there is a rather uniform growth of the red
function by $\sim 0.2\,$dex at all masses, accompanied by a comparable
uniform decrease in the blue function over the same mass range.  
Above $M_{\rm star}^{\rm crit}$, the red function grows by a larger factor
while the blue function remains roughly constant.  

Our earlier analysis can help us identify the processes responsible 
for the buildup of the red sequence at the low- and high-mass ends.
Red galaxies with $M_{\rm star}>M_{\rm star}^{\rm crit}$ consist of stars 
that have formed at $z\gg 1$, so their progenitors were already on the red 
sequence at $z\sim 1$.
These galaxies have continued assembling through dry mergers of red galaxies,
such that the growth of the red population at the high-mass end
comes largely at the expense of the red population at the low-mass end,
from below $M_{\rm star}^{\rm crit}$.
The true influx of galaxies into the low-mass red population 
equals the increase of the red mass function seen in 
Fig.~\ref{massfunction_evolution} plus the transfer of red galaxies
to the high-mass end by dry mergers.
This influx must come from blue galaxies migrating to the red sequence 
because the increase in total stellar mass in red galaxies
cannot come from new star formation in the red sequence
 \citep{faber_etal07}.

Part of the stellar mass transferred from the blue to the red sequence 
is readily accounted for by the decrease
in the number of blue galaxies at  $M_{\rm star}<M_{\rm star}^{\rm crit}$, 
where most of the blue sequence mass resides.
However, the total stellar mass in galaxies with $M_{\rm star}>10^{10}M_\odot$
has increased from $z\sim 1$ to $z\sim 0$. This implies that, despite the 
drop in the number of blue galaxies, new stellar mass must have been formed
along the blue sequence at $0\lsim z\lsim 1$.
In our model, this is mainly due to the conversion of disc gas into stars. 
Only to a lesser extent is it due to new galaxies moving into the blue 
sequence due to growth through the accretion of gas and wet mergers.

We have thus identified three fluxes that drive the evolution of 
the populations of red and blue galaxies in the colour-mass diagram
from $z\sim 1$ to $z\sim 0$. 
First, the flux of stellar mass into the blue sequence due to conversion of 
gas into stars in blue galaxies $\lsim M_{\rm star}^{\rm crit}$.
Second, the transition of blue galaxies onto the red sequence at 
$\lsim M_{\rm star}^{\rm crit}$.
Third, the transfer of red galaxies by dry mergers from the low-mass part 
of the red sequence to the high-mass part, 
$M_{\rm star}> M_{\rm star}^{\rm crit}$. 
Only the first mode generates new stellar mass in the
overall galaxy population, and only the first two contribute to the growth 
of stellar mass in the red sequence.

Now, with the benefit of hindsight, we are ready to interpret the
changes seen in Fig.~\ref{massfunction_evolution}.
One obvious trend in both the model and the figure is the assembly upsizing 
that creates the most massive elliptical galaxies on the red sequence at 
late times.  This is visible in figure as the rightward 
shift in the red function at high masses. 
These galaxies have assembled late and therefore appear at
high masses late.  Less clear, however, is the behaviour of the red mass function below
$M_{\rm star}^{\rm crit}$.  We know from the detailed histories of 
galaxies in Fig.~\ref{times} that the star-formation downsizing trend at 
these masses is strong, yet this effect is not obvious in the mass functions.  
All masses below $M_{\rm star}^{\rm crit}$ appear to grow about the same. 
To be precise, there does not
appear to be much greater growth at the lowest masses on the red
sequence compared to the growth at $M_{\rm star}^{\rm crit}$.
This happens largely because our model contain many satellite disc 
galaxies that form too early and are quenched too early
(Fig.~\ref{massfunctions} in the Appendix).
The lesson is that deducing either down- or upsizing trends from 
the mass function is tricky.

Fig.~\ref{massfunction_evolution} also sheds light on a common 
misinterpretation of the crossover mass, that is, 
the point where the red function intersects the blue function.
Downward movement of this point in time has been suggested as an
indicator of downsizing \citep{bundy_etal06,hopkins_etal07}.
Even though downsizing is actually taking place at these
masses in the model, we have seen that this is not evident from each of the
mass functions alone.  Instead, Fig.~\ref{massfunction_evolution} shows 
quite clearly that the crossover point is moving downward simply because 
the number of red galaxies is growing relative to the number of blue galaxies,
independent of how these changes depend on mass.
This effect is enhanced when the mass functions are relatively flat,
which is true both in the model and in real data (see the Appendix).  
The crossover point would move downward even when blue galaxies become
red in a uniform rate at all masses, namely without any real downsizing trend
in the SFR along the blue cloud or in the buildup of the red sequence. 

In the Appendix, we make three points concerning the observed mass functions.
First, determining the number of very massive galaxies at the steep bright
end is tricky because even small random mass errors create large errors 
in the counts.  The observed mass function
at the bright end is not the true function but rather the true
function smoothed by errors.  Owing to the steepness of the bright
end, this tends to falsely inflate the number of massive objects (the
so-called Eddington effect; see also \citealp{kitzbichler_white06} 
and \citealp{lauer_etal07}).
The Appendix shows that random rms errors of only
$0.2\,$dex in mass inflate the observed counts by a full order of magnitude
at $M_{\rm star}\sim 10^{11.5}M_\odot$.  Model counts must therefore be 
smoothed to mimic this effect before comparing to data, but it is unclear 
whether the errors get smaller or larger with redshift, and thus whether 
the evolutionary effect is positive or negative.

Second, the total magnitudes of massive ellipticals may be systematically
underestimated because the light profiles of very massive galaxies are nearly
isothermal and it is hard to estimate the total light, let alone
distinguish galaxy light from intracluster light.  This problem likely
worsens at low redshifts, and so the sense is probably to
underestimate galaxy masses, and thus the number of massive galaxies
towards low redshifts.
 
Third, before comparing mass functions from different authors, it is
necessary to convert them to the same assumed stellar initial mass 
function (IMF).  Certain
authors who formerly appeared to agree now disagree after this
correction is made. Consistent trends in even just the raw counts
are not yet evident.

Given these difficulties, both observational and theoretical, we do
not attempt in this paper to deduce either down- or upsizing trends
from luminosity functions or mass functions.  All that can be 
said is that the number of red galaxies has increased by at least a
factor of 2 near $L_*$ and $M_*$ since $z = 1$, while the number of blue
galaxies has not increased since $z=1$ \citep{faber_etal07}.  These
trends do not alone argue for down- or upsizing, but at least they are 
in reasonable agreement with the results of our model.

\section{Discussion and conclusion}
\label{sec:conc}

We showed that the downsizing of elliptical galaxies, where the more
massive galaxies formed their stars earlier and over a shorter period,
is not in conflict with the standard hierarchical clustering scenario.
We demonstrated that this is in fact a natural outcome of 
the shutdown of star formation in haloes above a critical mass $M_{\rm crit} \sim 10^{12} M_\odot$.
Such a shutdown is essential for explaining the distribution of galaxies 
in luminosity and colour \citep{cattaneo_etal06,bower_etal06,croton_etal06}.
In particular, this shutdown prevents the overgrowth of the central galaxies 
in massive haloes beyond the observed upper limit for bright blue galaxies,
and allows the appearance of massive red and dead galaxies, thus reproducing
the observed galaxy bimodality.
The critical mass for shutdown originates from stable virial shock heating, 
which occurs in haloes above this threshold and shuts down the cold gas 
supply for star formation there 
\citep{birnboim_dekel03,binney04,keres_etal05,dekel_birnboim06}.
This scenario only works if gas in massive haloes is maintained hot by a
continuous energy source that balances the radiative cooling over long 
cosmological periods.
The mechanical energy of jets from low-power radio AGNs is a promising
candidate  \citep{fabian_etal03,ruszkowski_etal04,forman_etal05,best_etal05,
voit_donahue05,best_etal06,dunn_fabian06},  
which can result in self-regulated accretion and AGN activity capable
of maintaining the required long-term shutdown 
\citep{omma_binney04,rafferty_etal06,cattaneo_teyssier07}.
An alternative/additional source of long-term quenching is the gravitational energy
of accretion into the centres of the potential wells of dark matter haloes 
\citep{birnboim_etal07,dekel_birnboim07,khochfar_ostriker07}.

The shutdown at a critical halo mass introduces an entry mass 
$M_{\rm star}^{\rm crit}$ to the red sequence.
Central galaxies cease to make stars and turn red and dead at a mass equal to
the characteristic mass of the central galaxy in a halo of mass $M_{\rm crit}$.
The central galaxies of the most massive haloes reach 
$M_{\rm star}^{\rm crit}\sim 10^{11}M_\odot$ earlier 
(Fig.~\ref{halo_and_galaxy_growth}).
Therefore, they have more time to grow above $M_{\rm star}^{\rm crit}$ 
by dry merging along the red sequence. 
For this reason, the central galaxies of the most massive haloes are the 
most massive galaxies in the Universe and those that contain the oldest 
stellar populations.
In our model, the early growth of the central galaxies of the most massive 
haloes is also helped by the fact that $M_{\rm crit}$ is assumed to be
higher than $2\times 10^{12} M_\odot$ before $z\gsim 3$, based on the
predictions of \citet{dekel_birnboim06}.

We demonstrated that shutdown-driven downsizing 
is in agreement with the observed archaeological downsizing of 
elliptical galaxies, as measured by Thomas et al. (2005; Fig.~\ref{agevsmass}).
The introduction of shutdown above a threshold mass also explains why the
emergence of the red sequence has started already at $z\sim 2$ 
and why the shutdown begins at high masses (Fig.~\ref{bimodality_emergence}).
Over $3/4$ of the model ellipticals with $M_{\rm star}>3\times10^{11}M_\odot$
were already on the red sequence by $z\sim 2$, in agreement with
FIRES observations \citep{giallongo_etal05}.
In contrast, the buildup of the red sequence at 
$M_{\rm star}\lsim 10^{11}M_\odot$ continues at $z\lsim 1$.

The downsizing in star formation and the upsizing in  mass assembly
affect the evolution of the mass functions of red and blue galaxies.
The three phenomena that control the evolution of the mass functions are:
the formation of new stellar mass in blue sequence galaxies, 
the migration of blue galaxies onto the 
red sequence, and the migration of red galaxies from the low-mass end to the high-mass end of the red sequence due to dry merging.
However, the combination of downsizing in star formation and upsizing in  mass assembly is such that the evolution of the mass functions 
is a poor indicator of downsizing even if the data were perfectly accurate (the large uncertainties that still exist in the data themselves have
been discussed separately in Appendix~A).

Giant ellipticals grow by $\sim 0.4-0.5\,$dex after entering the red 
sequence and by $\sim 0.2-0.3\,$dex between $z\sim 1$ and $z\sim 0$. 
Within the red population, they are those that have finished to assemble 
their mass most recently 
(Fig.~\ref{times}e, red symbols; also \citealp{delucia_etal06} and
\citealp{delucia_blaizot07}), despite the fact that their final mass was 
turned into stars earlier than in other galaxies (Fig.~\ref{times}e, 
purple symbols),
due to the early shutdown of star formation in these objects.

In this paper we have focussed on the contribution of the shutdown above a critical
halo mass to the downsizing.
In principle,there could be other reasons for the observed downsizing.
Nevertheless, the comparison of the stellar ages in the two models,
which differ only by imposing an explicit shutdown,
provides clean evidence that a shutdown of this sort
can be the main driver of the observed downsizing.


\section*{Acknowledgments}
We acknowledge stimulating discussions with J.~Devriendt, M.~Pannella 
and R.~Somerville.
The simulations presented in this article were run on the Horizon 
supercomputer at the CRAL in Lyon within the framework of the Horizon Project. 
This research has been partly supported by ISF 213/02, by GIF I-895-207.7/2005,
by a France-Israel grant, by the Einstein Center at HU, 
and by NASA ATP NAG5-8218.


\bibliographystyle{mn2e}

\bibliography{ref_av}

\appendix

\section{Galaxy mass functions at  $z\sim 0$ and $z\sim 1$}

The evolution of the mass functions of red and blue galaxies,
whether increasing more strongly at low mass or at high mass, has
been taken as a barometer of whether the red sequence forms primarily 
through downsizing or upsizing (\citealp{bundy_etal05,borch_etal06,pannella_etal06}; also see the discussion
in \citealp{faber_etal07}).
In Section~7, we have shown how the evolution of the mass function of red galaxies is driven by a combination of 
star formation downsizing (due to quenching)
and stellar mass assembly upsizing (due to merging).  
We have argued that the inherent difficulties in interpreting model mass functions, together with the large uncertainties that are still present in the models themselves, should suggest caution in using
mass function data to infer up- or downsizing. In this Appendix, we consider the observational data on mass functions at $z\sim 0$ and $z\sim 1$.

The SDSS and 2MASS are the main data sets used to study the mass function of galaxies at $z\sim 0$ (e.g. \citealp{bell_etal03}).
Meanwhile, surveys that are increasingly deep (MUNICS, FDF, GOODS) are beginning to give us estimates for the mass function of galaxies at $z\sim 1$ (e.g. \citealp{bundy_etal06,pannella_etal06}). 
Observational determinations of galaxy mass functions are affected by three sources of error, which we have enumerated in Section~7, and which we now discuss in greater detail.

\begin{figure*}
\noindent
\label{mhalo_mgal}
\begin{minipage}{8.4cm}
  \centerline{\hbox{
      \psfig{figure=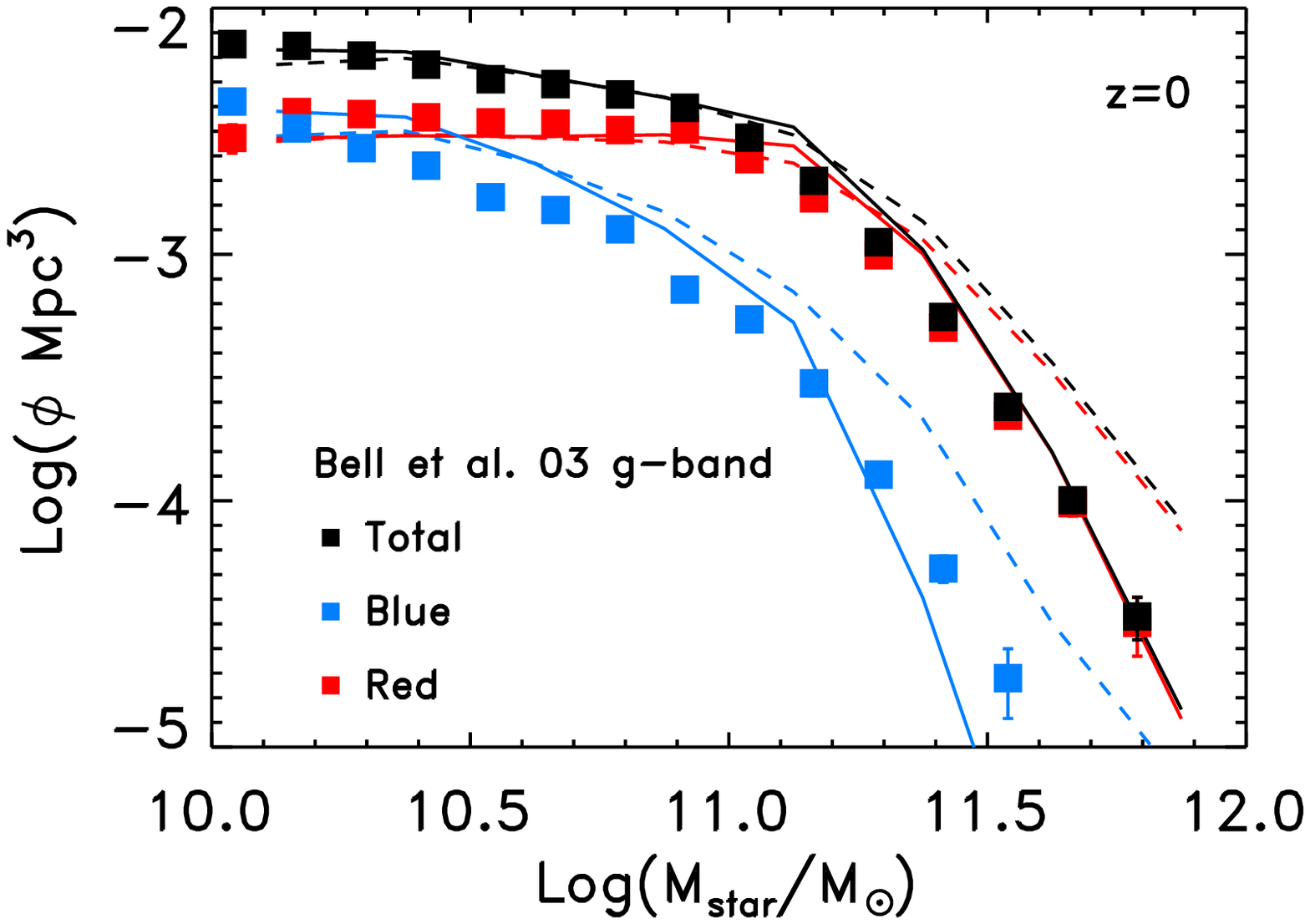,height=6.43cm,angle=0}
  }}
\end{minipage}\    \
\begin{minipage}{8.4cm}
  \centerline{\hbox{
      \psfig{figure=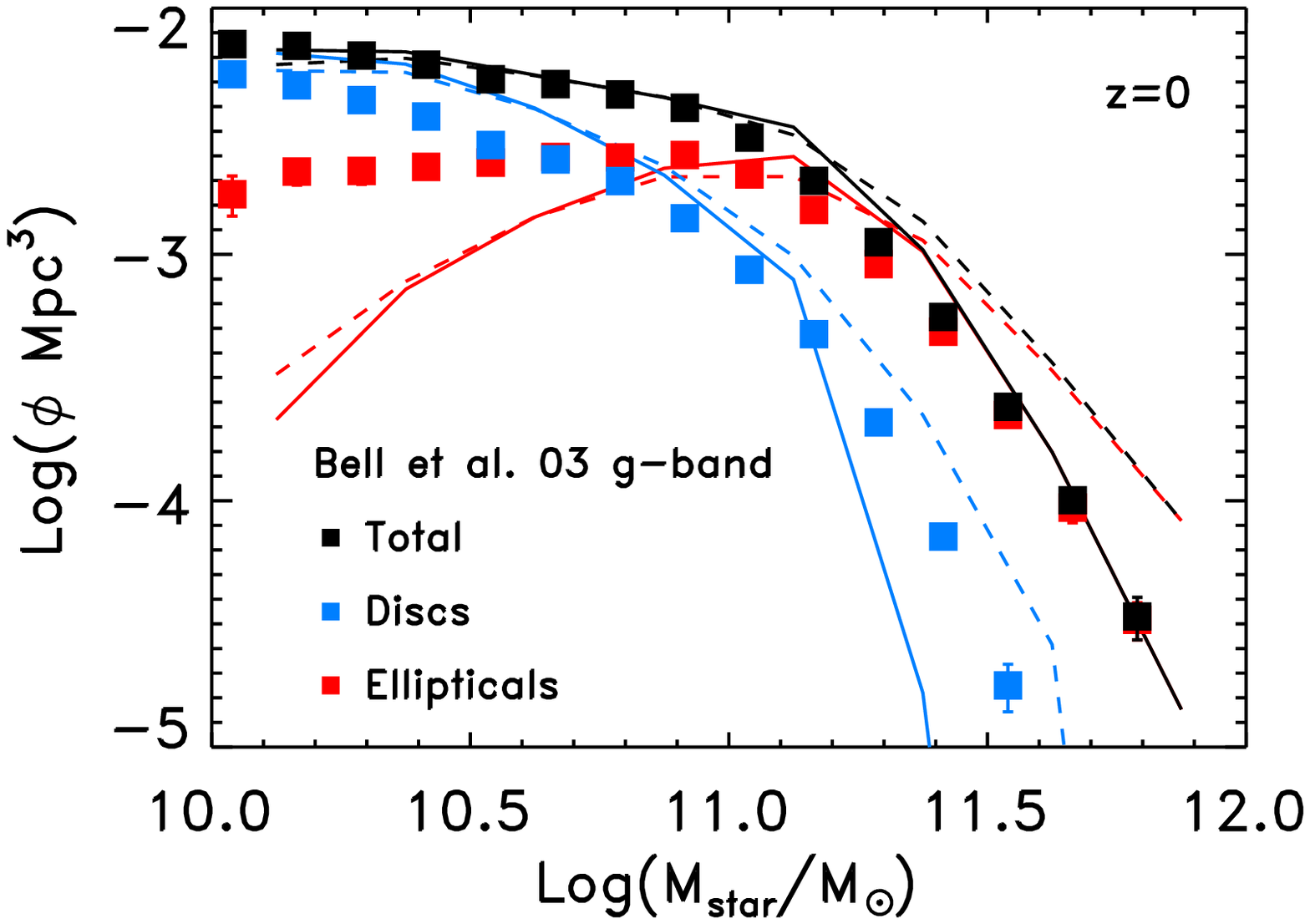,height=6.43cm,angle=0}
  }}
\end{minipage}\    \
\begin{minipage}{8.4cm}
  \centerline{\hbox{
      \psfig{figure=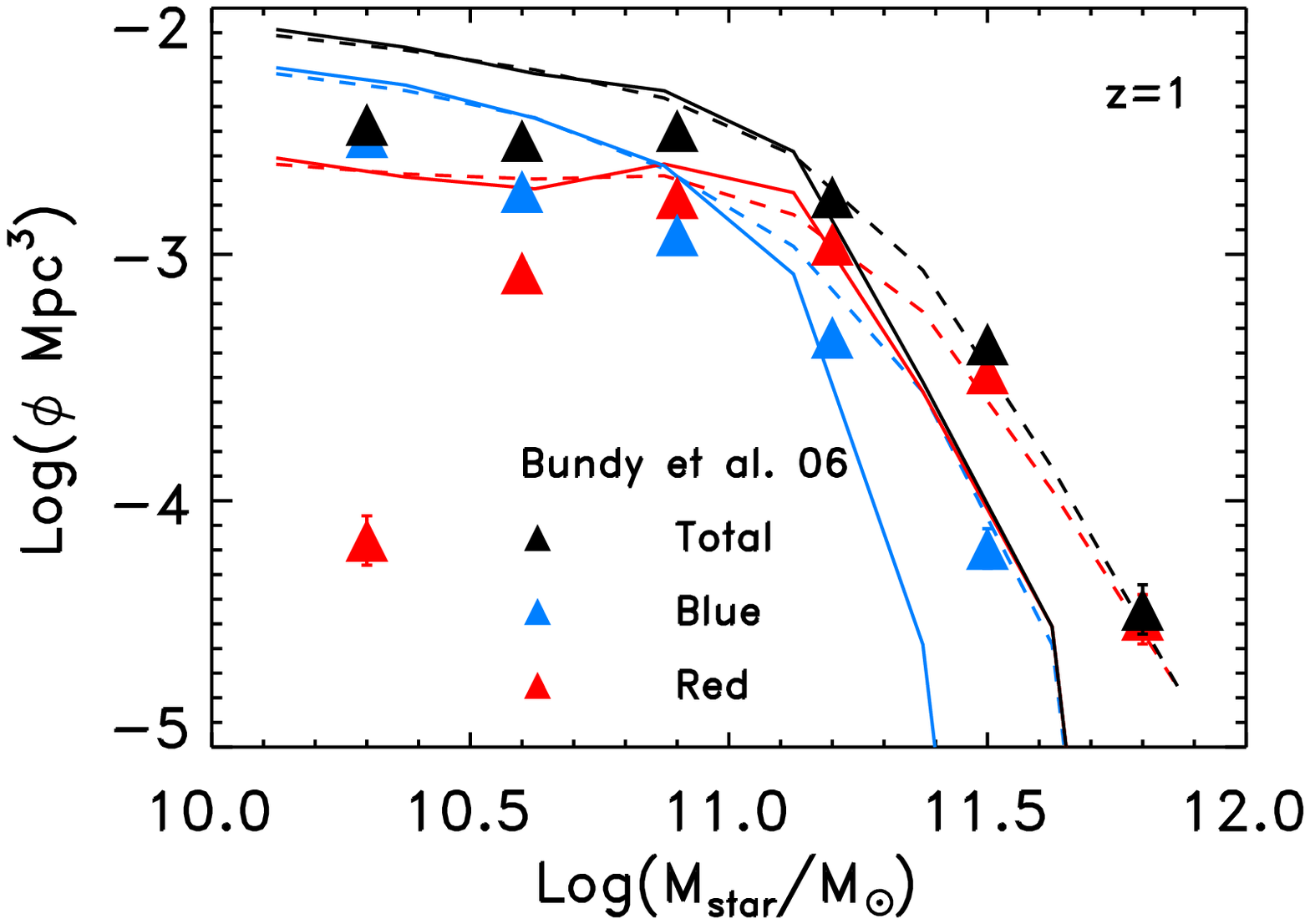,height=6.43cm,angle=0}
  }}
\end{minipage}\    \
\begin{minipage}{8.4cm}
  \centerline{\hbox{
      \psfig{figure=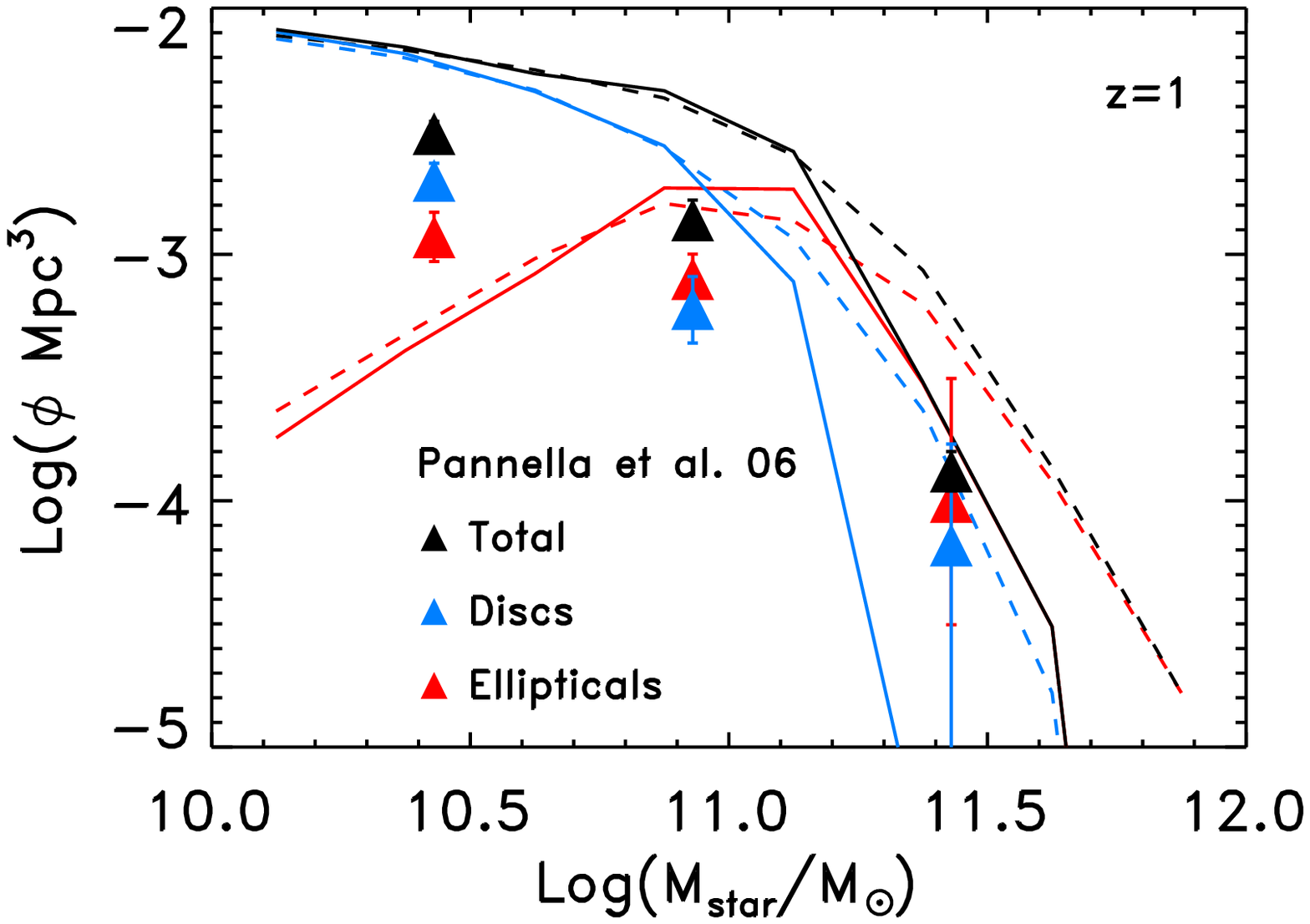,height=6.43cm,angle=0}
  }}
\end{minipage}\    \
\caption{Mass functions at $z=0$ and $z=1$ for early type galaxies (red) and late type galaxies (blue) classified spectrally (left) and morphologically (right)
The coloured and black points with error bars are the data.
The lines show the predictions of our new model with star formation shutdown without (solid lines) and with (dashed lines) the inclusion of a 0.2$\,$dex Gaussian random error. Note how the predictions of the model are completely altered at the massive end when such smoothing is included.}
 \label{massfunctions}
\end{figure*}

The first is the steep slope of the mass function at the bright
end.
Therefore, even small random mass
errors create large errors in the counts.  The observed mass function
at the bright end is not the true function but rather the true
function smoothed by errors.  Owing to the steepness of the bright
end, this tends to falsely inflate the number of massive objects (the
so-called Eddington effect; see also \citealp{kitzbichler_white06} and \citealp{lauer_etal07}).
Random rms errors of only
$0.2\,$dex in mass inflate observed counts by a full order of magnitude
at $M_{\rm star}\sim 10^{11.5}M_\odot$ (compare the solid and the dashed lines in Fig.~\ref{massfunctions}).  
Model counts must therefore be smoothed to mimic
this effect before comparing to data, but it is unclear whether errors
get smaller or larger with redshift, and thus whether the evolutionary
effect is positive or negative.
Fig.~\ref{massfunctions} shows that we can make our model agree with one data set or the other simply by 
making different assumptions concerning the amplitude of the random measurement errors.

A second source that affects the photometry of the most massive galaxies is their having shallow light profiles that do not converge to a total magnitude.
The magnitudes of very bright ellipticals are hard to measure
on account of their extended envelopes, which may be growing with time due to galaxy-galaxy interactions.
It is possible that many of them extend outside the photometric aperture used to measure their luminosities.
Simply fitting such galaxies with de Vaucouleurs $R^{1/4}$ profiles is not appropriate because many profiles (at least locally) are even more extended than that (e.g., \citealp{graham_etal01}). Total magnitude errors may therefore be larger for nearby galaxies than for distant ones. For example, the total luminosities of the brightest cluster galaxies of nearby Abell clusters can vary by many tenths of a magnitude \citep{gonzalez_etal05} depending on the profile fitting method used.
\citet{lauer_etal07} find errors of up to a magnitude in SDSS luminosities for such galaxies. These are very large errors when one considers that a change of only 0.2$\,$mag at $4\,L_*$ translates to a change in number density of a factor of two in the Schechter function. Stellar masses have even larger errors because they depend on assumed evolution models for the fade in $M/L_B$, which introduce an additional uncertainty of a few tenths of a magnitude. 
While random errors are more likely to be important at high redshift, the problem deriving from the extended envelops of massive ellipticals is likely to worsen at low redshift, since a lot of the growth by dry merging occurs between $z\sim 1$ and $z\sim 0$ (Fig.~\ref{halo_and_galaxy_growth}).
Therefore, the combination of these two sources of error is more likely to
underestimate  the evolution of the number of massive early type galaxies from $z\sim 1$ to $z\sim 0$.

Thirdly, before comparing mass functions from different authors, it is
necessary to convert them first to the same assumed IMF.  
\citet{bell_etal03} used a `diet' Salpeter IMF,
which corresponds to a Salpeter IMF truncated at low stellar masses \citep{bell_dejong01}.
To convert the masses obtained  \citet{pannella_etal06} for a standard Salpeter IMF into masses that can be
compared with the results by \citet{bell_etal03}, one must subtract $\sim 0.15\,$dex from the masses determined by \citet{pannella_etal06}. \citet{bundy_etal06} used a Chabrier IMF. The masses estimated with the Chabrier IMF are lower than those estimated with a Salpeter IMF by $\sim 0.25\,$dex. Therefore, we must add $0.1\,$dex to the masses determined by \citet{bundy_etal06} to compare their mass function with that by \citet{bell_etal03}.
After performing these corrections, we see that the apparent agreement between
the figures in \citet{bundy_etal06} and \citet{pannella_etal06}
 is simply due to the fact that they used different IMFs.
For reasons that we cannot easily  interpret, with the same IMF, the galaxy mass function inferred by \citet{bundy_etal06} contains many more massive galaxies at $z\sim 1$ than the mass function inferred
by Pannella et al. (2006; Fig.~\ref{massfunctions}).
The data by \citet{bundy_etal06} are consistent with no evolution at $M_{\rm star}\gsim 10^{11}M_\odot$
in the redshift interval $0\lsim z\lsim 1$.
The data by \citet{pannella_etal06} are consistent with growth in mass by $0.2-03\,$dex at 
$M_{\rm star}\gsim 10^{11}M_\odot$

In conclusion, it appears that the present state of mass functions
    at $z = 0$ and $z = 1$ is sufficiently confused to prevent firm
    conclusions.  Problems are particularly severe at the high-mass
    end, where random errors seriously inflate counts at a given mass and 
    questions persist about how to measure total light and mass.  
    Fortunately, uncertainties are much smaller near $L_*$.
    Though the data are not perfect, the general trend is for
    red galaxies to increase in number with time near the knee of the
    mass function, by amounts that are in broad agreement
    with both luminosity function data \citep{faber_etal07} and with the predictions of
    our model.

\label{lastpage}
\end{document}